# UNIVERSAL DENSITY MATRIX FOR THE PHASE SPACE

**E.E. Perepelkin[a], B.I. Sadovnikov[a], N.G. Inozemtseva[b], E.V. Burlakov[a]**

[a] *Faculty of Physics, Lomonosov Moscow State University, Moscow, 119991 Russia*
*E. Perepelkin e-mail: pevgeny@jinr.ru, B. Sadovnikov e-mail: sadovnikov@phys.msu.ru*
[b] *Dubna State University, Moscow region, Moscow,141980 Russia*
*e-mail: nginozv@mail.ru*

**Abstract**

In this paper, a new representation of the Wigner function for a quantum system in the phase space is proposed. The new representation is of the form $W = \mathrm{Sp}[\rho \mathcal{W}]$, where $\rho$ is the density matrix, and $\mathcal{W}$ is the universal density matrix. The density matrix $\rho$ for each quantum system is different, and the universal matrix $\mathcal{W}$ is the same for any quantum system.

Thus, the matrix $\mathcal{W}$ has a fundamental character. In the work, the elements of the universal matrix $\mathcal{W}$ were found explicitly and their properties were investigated. The diagonal elements of the matrix $\mathcal{W}$ are the Wigner functions of the harmonic oscillator, which do not introduce dissipation into the quantum system. The off-diagonal elements of the matrix $\mathcal{W}$ contain frequency oscillations responsible for dissipations in the quantum systems.

**Introduction**

When considering quantum systems, the coordinate $\Psi(\vec{r},t)$ and momentum $\tilde{\Psi}(\vec{p},t)$ representations of the wave function are used, which are connected by the Fourier transform $\mathcal{F}[\Psi] = \tilde{\Psi}$ [1-3]. Thus, the probability density function of the momentum $|\tilde{\Psi}(\vec{p},t)|^2$ is constructed from the probability density function of the coordinate $|\Psi(\vec{r},t)|^2$, since $|\tilde{\Psi}(\vec{p},t)|^2 = |\mathcal{F}[\Psi(\vec{r},t)]|^2$. Consideration of quantum systems in the phase space $(\vec{r}, \vec{p})$ was first made by E.P. Wigner, H. Weyl [4-6]. The Wigner function $W(\vec{r}, \vec{p})$ determines the density of quasi-probabilities of the random variables $\vec{R}$ and $\vec{P}$ for a quantum system in the phase space [7, 8, 30–34]. The construction of the Wigner function was done phenomenologically using the Fourier transform. In addition to the quasi-probability function $W(\vec{r}, \vec{p})$, phenomenologically constructed probability $Q, P$ – functions, parameterization are used, which are positive [9–13]. Despite the phenomenological nature of such constructions, the Weyl-Wigner-Moyal-Groenewold formalism is widely used when considering quantum systems in the phase space [14-16, 27].

For an arbitrary quantum system described by the wave function $\Psi$, the determination of the Wigner function requires the calculation of an integral of the form

$$W(x,p,t) = \frac{1}{2\pi\hbar} \int_{-\infty}^{+\infty} \exp\left(-i\frac{ps}{\hbar}\right) \left\langle x+\frac{s}{2} \middle| \hat{\rho}(t) \middle| x-\frac{s}{2} \right\rangle ds, \qquad (\text{i.1})$$

where $\hat{\rho}$ is the density matrix. In the general case, the calculation of the integral (i.1) is analytically difficult, and in most cases, numerical integration is performed. Knowing the exact expression of the integral (i.1) allows analytic analysis of the behavior of complex dissipative quantum systems.

The form of the Wigner function (i.1) depends on the wave function $\Psi$ of the quantum system, which may have different representations. In this work, to obtain a single representation of the Wigner function $W$ for an arbitrary wave function $\Psi$, we expand the wave function $\Psi$ in a series in terms of the basic wave functions $\Psi_n$ of a quantum harmonic oscillator. The problem for a quantum harmonic oscillator in the phase space is well studied, and has an expression for the Wigner function.

This approach allows one to represent the Wigner function for an arbitrary quantum system in the form

$$W = \mathrm{Sp}[\hat{\rho}\mathcal{W}], \qquad (i.2)$$

where $\mathcal{W}$ is a matrix, which we call in this paper a universal density matrix. The term «universal» is associated with the fact that the matrix $\mathcal{W}$ has a single structure for any quantum system. Therefore, knowing the density matrix $\hat{\rho}$ using the formula (i.2), one can get a representation for the Wigner function of an arbitrary quantum system.

The universal density matrix $\mathcal{W}$ has the following structure. On the main diagonal of the matrix $\mathcal{W}$, the well-known Wigner functions $w_{n,n}$ for the harmonic oscillator are located, which correspond to different energy levels of the oscillator $\langle\langle \varepsilon_n \rangle\rangle$. As is known, such functions are valid, but having negative values. When moving along concentric circles in the phase plane (at a constant energy $\varepsilon = const$), the diagonal elements of the matrix $\mathcal{W}$ have constant values $w_{n,n} = const$. The elements of the matrix $\mathcal{W}$ located on the upper and lower diagonals $w_{n,k}$ ($n \neq k$) are complex functions and in this work they are found explicitly. The matrix elements $w_{n,k}$ have a frequency of oscillations $|\varpi_{n,k}| = |n-k|$, which grows with the distance from the main diagonal. Thus, when moving along the phase trajectories $\varepsilon = const$, the functions $w_{n,k}$ oscillate with frequency $|\varpi_{n,k}|$. The values of the functions $w_{n,k}$ belong to a multivalent Riemann surface. The frequency of oscillations $|\varpi_{n,k}|$ corresponds to the valent number of the Riemann surface.

As a result, an arbitrary quantum system is represented as an infinite set of harmonic oscillators. The Wigner function of such a system is represented as the superposition of the probability density functions $w_{n,k}$. When a quantum system moves along the phase trajectories $\mathcal{E} = \frac{p^2}{2m} + U = const$, the probability density $W$ changes. Such a change is due to the presence of oscillations in the elements $w_{n,k}$ ($n \neq k$). This fact is well known when comparing classical and quantum systems with potentials $\frac{\partial^{2l+1}U}{\partial x^{2l+1}} \neq 0$ at $l > 0$. The Moyal [26] and Liouville equations coincide for potentials $\frac{\partial^{2l+1}U}{\partial x^{2l+1}} = 0$ at $l > 0$ and the probability density is constant $W = const$ along the phase trajectory $\mathcal{E} = const$. The presence of higher derivatives $\frac{\partial^{2l+1}U}{\partial x^{2l+1}} \neq 0$ at $l > 0$ in the right side of the Moyal equation leads to dissipative processes that change the probability density $W \neq const$ along the phase trajectory $\mathcal{E} = const$. This paper shows that such changes may have a periodic structure with oscillation frequencies $|\varpi_{n,k}|$.

The paper has the following structure. In §1, the explicit form of the matrix elements $w_{n,k}$ of the universal density matrix $\mathcal{W}$ is obtained. The matrix elements $w_{n,k}$ are representable in the form $w_{n,k}(x,p) \sim e^{-2\varepsilon}\mathcal{P}_{n,k}$, where $\mathcal{P}_{n,k}(z_1, z_2)$, $z_1, z_2 \in \mathbb{C}$ are new polynomials. In §1, a series of theorems on the properties of polynomials $\mathcal{P}_{n,k}$ is proved. The orthogonality of the polynomials $\mathcal{P}_{n,k}$ is investigated. At $n = k$ the $\mathcal{P}_{n,k}$ are transformed into the famous Laguerre polynomials $\mathcal{P}_{n,n}(\varsigma_1, \varsigma_2) = L_n(-2\varsigma_1\varsigma_2)$. In §1, the representation of the Wigner function is obtained in the form

$$W(x,p) = \frac{e^{-2\varepsilon(x,p)}}{\pi\hbar} \sum_{n,k=0}^{+\infty} |\rho_{k,n}| \Upsilon_{n,k}\left(\sqrt{2\varepsilon(x,p)}\right) \vec{n}_k^T \Omega^{(n,k)}(\varphi) \vec{n}_n, \qquad (\text{i.3})$$

where $\Upsilon_{n,k}$ is polynomials, associated with the polynomials $\mathcal{P}_{n,k}$; $\Omega^{(n,k)}(\varphi)$ is a rotation matrix in the phase plane; $\vec{n}_k$ are unit vectors associated with the density matrix $\rho$; the angle $\varphi$ corresponds to a point $(x,p)$ on the phase plane.

In §2, we consider the connection between the Moyal equation and the Vlasov equation for classical and quantum dissipative systems. In §2, an extended approximation of the mean acceleration flow $\langle \dot{\vec{v}} \rangle$ is introduced for the Vlasov equation, which was called the Vlasov-Moyal approximation

$$\langle \dot{v}_\alpha \rangle = \sum_{n=0}^{+\infty} \frac{(-1)^{n+1}(\hbar/2)^{2n}}{m^{2n+1}(2n+1)!} \frac{\partial^{2n+1} U}{\partial x_\alpha^{2n+1}} \frac{1}{f_2} \frac{\partial^{2n} f_2}{\partial v_\alpha^{2n}}, \qquad (\text{i.4})$$

where $f_2 = f_2(\vec{r}, \vec{v}, t)$ is the probability density function from the Vlasov equation [20-23]. The Vlasov equation is a kinematic equation in which there is a quantity $\langle \dot{\vec{v}} \rangle$

$$\frac{\partial f_2}{\partial t} + \operatorname{div}_r\left[\vec{v} f_2\right] + \operatorname{div}_v\left[\langle \dot{\vec{v}} \rangle f_2\right] = 0. \qquad (\text{i.5})$$

When using the approximation (i.4), the equation (i.5) goes into the Moyal equation. In §2, expressions for dissipation sources $\langle Q_2 \rangle$, $\langle\langle Q_2 \rangle\rangle$ ($Q_2 = \operatorname{div}_v \langle \dot{\vec{v}} \rangle$), which determine the behavior of the Boltzmann $H_2$- function (the entropy of the system) [25], are obtained.

In §3, under the condition that the potential energy $U$ is expanded into a power series, an expression is obtained for the energy $\langle\langle \mathcal{E} \rangle\rangle$ of the quantum system

$$\langle\langle \mathcal{E} \rangle\rangle = \int_{-\infty}^{+\infty}\int_{-\infty}^{+\infty} \mathcal{E}(x,p) W(x,p) dx dp, \qquad (\text{i.6})$$

$$\langle\langle \mathcal{E} \rangle\rangle = \sum_{n=0}^{+\infty} |\rho_{n,n}| \langle\langle \mathcal{E}_n \rangle\rangle +$$

$$+\sum_{n,k=0}^{+\infty}|\rho_{k,n}|\cos(\alpha_k-\alpha_n)\sqrt{2^{n+k}n!k!}\sum_{\frac{n-k+l}{2}\in\mathbb{Z},\,l\geq|n-k|}a_l\left(\frac{\hbar}{4m\omega}\right)^{l/2}C_l^{\frac{n-k+l}{2}}\sum_{s=0}^{\min(n,k)}\frac{(-1)^s\left(\frac{k+n+l}{2}-s\right)!}{2^s s!(k-s)!(n-s)!},$$

where $\rho_{k,n}=c_k\bar{c}_n$ are the matrix elements of the density matrix; $\alpha_k=\arg c_k$; $C_n^k=\dfrac{n!}{k!(n-k)!}$ is the number of combinations; $a_l$ are serial expansion coefficients of the potential energy; $\langle\langle\varepsilon_n\rangle\rangle=\hbar\omega\left(n+\dfrac{1}{2}\right)$ are energy levels of the harmonic oscillator.

In conclusion, a brief list of the main results of the work.

## §1 Universal density matrix

The solutions $\Psi_n$ for a quantum harmonic oscillator form an orthonormal basis in the space $L_2$ [18,19]

$$\Psi_n(x)=\frac{1}{\sqrt{2^n n!}}\left(\frac{m\omega}{\pi\hbar}\right)^{\frac{1}{4}}e^{-\frac{m\omega x^2}{2\hbar}}H_n\left(\sqrt{\frac{m\omega}{\hbar}}x\right),\;n\in\mathbb{N}_0,\qquad(1.1)$$

$$\tilde{\Psi}_n(p)=\frac{(-i)^n}{\sqrt{2^n n!}}\left(\frac{m\omega}{\pi\hbar}\right)^{\frac{1}{4}}\frac{1}{\sqrt{m\omega}}e^{-\frac{p^2}{2m\omega\hbar}}H_n\left(\frac{p}{\sqrt{m\omega\hbar}}\right),$$

where

$$\tilde{\Psi}_n(p)=\frac{1}{\sqrt{2\pi\hbar}}\int\Psi_n(x)e^{-i\frac{px}{\hbar}}dx,\;\Psi_n(x)=\frac{1}{\sqrt{2\pi\hbar}}\int\tilde{\Psi}_n(p)e^{i\frac{px}{\hbar}}dp,$$

where $H_n$ is Hermitian polynomials. Consequently, a certain solution $\Psi\in L_2$ of the Schrödinger equation can be expanded in the basis $\{\Psi_n\}$

$$\Psi(x,t)=\sum_{n=0}^{+\infty}c_n(t)\Psi_n(x),\qquad\tilde{\Psi}(p,t)=\sum_{n=0}^{+\infty}c_n(t)\tilde{\Psi}_n(p),\qquad(1.2)$$

$$|\Psi(x,t)|^2=\sum_{n,k=0}^{+\infty}c_n\bar{c}_k\Psi_n\bar{\Psi}_k,\;\int_{-\infty}^{+\infty}|\Psi|^2dx=\sum_{n=0}^{+\infty}|c_n|^2=1,$$

$$|\tilde{\Psi}(p,t)|^2=\sum_{n,k=0}^{+\infty}c_n\bar{c}_k\tilde{\Psi}_n\bar{\tilde{\Psi}}_k,\;\int_{-\infty}^{+\infty}|\tilde{\Psi}|^2dp=\sum_{n=0}^{+\infty}|c_n|^2=1.$$

It follows from the expansion (1.2) that the wave function $\Psi\in L_2$ of a quantum system is represented as a superposition of the wave functions of an oscillator (1.1). Knowing the wave function (1.2), we can construct the Wigner function

$$W(x,p)=\frac{1}{2\pi\hbar}\int_{-\infty}^{+\infty}\Psi\left(x+\frac{s}{2}\right)\bar{\Psi}\left(x-\frac{s}{2}\right)\exp\left(-i\frac{ps}{\hbar}\right)ds,\qquad(1.3)$$

$$W(x,p)=\frac{1}{2\pi\hbar}\int_{-\infty}^{+\infty}\bar{\tilde{\Psi}}\left(p-\frac{\xi}{2}\right)\tilde{\Psi}\left(p+\frac{\xi}{2}\right)\exp\left(i\frac{x\xi}{\hbar}\right)d\xi.$$

In accordance with (1.2) and (1.3), we define new functions $w_{n,k}(x,p)$.

*Definition 1*

Let $\Psi_n(x)$, $n,k \in \mathbb{N}_0$ be the wave functions corresponding to the quantum harmonic oscillator, then the set of functions of the form:

$$\mathcal{W} \stackrel{det}{=} \{w_{n,k}(x,p)\} \stackrel{det}{=} \frac{1}{2\pi\hbar} \int_{-\infty}^{+\infty} \overline{\Psi}_n\left(x - \frac{s}{2}\right) \Psi_k\left(x + \frac{s}{2}\right) \exp\left(-i\frac{ps}{\hbar}\right) ds = \qquad (1.4)$$

$$= \frac{1}{2\pi\hbar} \int_{-\infty}^{+\infty} \overline{\tilde{\Psi}}_n\left(p - \frac{\xi}{2}\right) \tilde{\Psi}_k\left(p + \frac{\xi}{2}\right) \exp\left(i\frac{x\xi}{\hbar}\right) d\xi.$$

will be called the universal density matrix in the phase space.

*Property 1*

For the elements $w_{n,k}(x,p)$ of the universal matrix $\mathcal{W}$, the following relations are true:

$$\int_{-\infty}^{+\infty} w_{n,k}(x,p) dp = \overline{\Psi}_n(x)\Psi_k(x), \qquad \int_{-\infty}^{+\infty} w_{n,k}(x,p) dx = \overline{\tilde{\Psi}}_n(p)\tilde{\Psi}_k(p). \qquad (1.5)$$

*Proof of Property 1*

Indeed, in view of Definition 1, we obtain

$$\int_{-\infty}^{+\infty} w_{n,k}(x,p) dp = \frac{1}{2\pi\hbar} \int_{-\infty}^{+\infty} e^{-i\frac{ps}{\hbar}} dp \int_{-\infty}^{+\infty} \overline{\Psi}_n\left(x - \frac{s}{2}\right) \Psi_k\left(x + \frac{s}{2}\right) ds =$$

$$= \int_{-\infty}^{+\infty} \delta(s) \overline{\Psi}_n\left(x - \frac{s}{2}\right) \Psi_k\left(x + \frac{s}{2}\right) ds = \overline{\Psi}_n(x)\Psi_k(x),$$

$$\int_{-\infty}^{+\infty} w_{n,k}(x,p) dx = \frac{1}{2\pi\hbar} \int_{-\infty}^{+\infty} e^{i\frac{x\xi}{\hbar}} dx \int_{-\infty}^{+\infty} \overline{\tilde{\Psi}}_n\left(p - \frac{\xi}{2}\right) \tilde{\Psi}_k\left(p + \frac{\xi}{2}\right) d\xi =$$

$$= \int_{-\infty}^{+\infty} \delta(\xi) \overline{\tilde{\Psi}}_n\left(p - \frac{\xi}{2}\right) \tilde{\Psi}_k\left(p + \frac{\xi}{2}\right) d\xi = \overline{\tilde{\Psi}}_n(p)\tilde{\Psi}_k(p),$$

which was to be proved.

*Property 2*

The diagonal elements $w_{n,n}$ of the universal density matrix $\mathcal{W}$ coincide with the Wigner functions (1.3) for the eigenstates $\Psi_n(x)$ of the harmonic oscillator.

$$w_{n,n}(x,p) = \frac{(-1)^n}{\pi\hbar} e^{-2\varepsilon(x,p)} L_n(4\varepsilon(x,p)), \qquad (1.6)$$

$$\varepsilon(x,p) = \frac{1}{\hbar\omega}\left(\frac{p^2}{2m} + \frac{m\omega^2 x^2}{2}\right),$$

where $L_n$ is Laguerre polynomials.

*Property 3*

The universal density matrix $\mathcal{W}$ is a Hermitian matrix $\mathcal{W}^\dagger = \mathcal{W}$, that is,

$$w_{n,k}(x,p) = \bar{w}_{k,n}(x,p). \qquad (1.7)$$

*Proof of Property 3*

Indeed, it follows from the definition (1.4) that

$$\bar{w}_{k,n}(x,p) = \frac{1}{2\pi\hbar} \int_{-\infty}^{+\infty} \Psi_k\left(x - \frac{s}{2}\right)\bar{\Psi}_n\left(x + \frac{s}{2}\right)\exp\left(i\frac{ps}{\hbar}\right)ds,$$

Let us replace the variables $s = -\xi$, $ds = -d\xi$. The limits of integration will be: $\xi_1 = -s_1 = +\infty$, $\xi_2 = -s_2 = -\infty$, therefore

$$\bar{w}_{k,n}(x,p) = -\frac{1}{2\pi\hbar}\int_{+\infty}^{-\infty} \bar{\Psi}_n\left(x - \frac{\xi}{2}\right)\Psi_k\left(x + \frac{\xi}{2}\right)\exp\left(-i\frac{p\xi}{\hbar}\right)d\xi =$$

$$= \frac{1}{2\pi\hbar}\int_{-\infty}^{+\infty} \bar{\Psi}_n\left(x - \frac{\xi}{2}\right)\Psi_k\left(x + \frac{\xi}{2}\right)\exp\left(-i\frac{p\xi}{\hbar}\right)d\xi = w_{n,k},$$

which was to be proved.

Using Definition 1, we write the Wigner function (1.3) of an arbitrary quantum system described by the wave function $\Psi$:

$$W(x,p) = \frac{1}{2\pi\hbar}\sum_{n,k=0}^{+\infty} c_k \bar{c}_n \int \bar{\Psi}_n\left(x - \frac{s}{2}\right)\Psi_k\left(x + \frac{s}{2}\right)\exp\left(-i\frac{ps}{\hbar}\right)ds = \sum_{n,k=0}^{+\infty} c_k \bar{c}_n w_{n,k}(x,p), \qquad (1.8)$$

$$W(x,p) = \sum_{n,k=0}^{+\infty} \rho_{k,n} w_{n,k}(x,p) = \mathrm{Sp}\left[\rho\mathcal{W}(x,p)\right],$$

$$\rho_{k,n} \stackrel{\text{det}}{=} c_k \bar{c}_n,$$

where $\rho_{k,n}$ will be called the density matrix. The density matrix $\rho$ is a Hermitian one, that is $\rho = \rho^\dagger$. According to the expressions (1.2), the condition $\mathrm{Sp}[\rho] = 1$ is satisfied for the density matrix $\rho$. It follows from the expression (1.8) that, knowing the universal density matrix $\mathcal{W}$, one can obtain the Wigner function of an arbitrary quantum system. In the matrix form, the expression (1.8) can be represented as a convolution

$$W = \bar{C}^T \mathcal{W} C, \qquad (1.9)$$

$$\bar{C}^T = \{\bar{c}_1, \bar{c}_2, ...\}, \quad C = \{c_1, c_2, ...\}^T.$$

Knowing the wave function $\Psi$, in accordance with (1.2) using the formula $c_n = \int_{-\infty}^{+\infty} \bar{\Psi}_n(x)\Psi(x)dx$, one can find the vector $C$ and use the formulas (1.9), (1.8) to obtain an expression for the Wigner function $W$ (1.3). On the other hand, the expression (1.8) in quantum mechanics is interpreted as the mean value of the operator $\mathcal{W}$. Consequently, according to (1.8), the Wigner function is the mean value of the operator $\mathcal{W}$.

The density matrix $\mathcal{W}$ will be universal for any quantum system described by the wave function $\Psi \in L_2$. Thus, the problem of constructing the Wigner function is reduced to finding a universal density matrix $\mathcal{W}$. Property 1 implies that it is necessary to find expressions for the functions $w_{n,k}(x, p)$ at $n \neq k$.

## Definition 2

Let $n, k \in \mathbb{N}_0$ and $z_1, z_2 \in \mathbb{C}$ and we define the binomial polynomials $\mathcal{P}_{n,k}(z_1, z_2)$ as

$$\mathcal{P}_{n,k}(z_1, z_2) \overset{\text{det}}{=} \frac{1}{\sqrt{2^{n+k} n! k!}} \sum_{s=0}^{\min(n,k)} \frac{1}{2^s s!} \frac{\partial^{2s}}{\partial z_1^s \partial z_2^s} \left[(2z_1)^n (2z_2)^k\right]. \quad (1.10)$$

## Corollary 1

It follows from the definition (1.10) that polynomials $\mathcal{P}_{n,k}(z_1, z_2)$ can be represented in the form

$$\mathcal{P}_{n,k}(z_1, z_2) = \sqrt{2^{n+k} n! k!} \sum_{s=0}^{\min(n,k)} \frac{z_1^{n-s} z_2^{k-s}}{2^s s!(k-s)!(n-s)!}. \quad (1.11)$$

and

$$\overline{\mathcal{P}}_{n+l,n}(-z, \overline{z}) = (-1)^l \mathcal{P}_{n,n+l}(-z, \overline{z}).$$

## Theorem 1

Let $n_1, k_1, n_2, k_2 \in \mathbb{N}_0$, $x, y \in \mathbb{R}$ and $\rho_2(x, y) = e^{-x^2 - y^2}$, then the integral

$$I = \int_{-\infty}^{+\infty} \int_{-\infty}^{+\infty} \rho_2(x, y) \mathcal{P}_{n_1, k_1}(x, y) \mathcal{P}_{n_2, k_2}(x, y) dx dy, \quad (1.12)$$

equals zero, i.e. $I = 0$ at $n_1 + n_2$ being even and $k_1 + k_2$ being odd or at $n_1 + n_2$ being odd and $k_1 + k_2$ being even.

At $n_1 + n_2$ and $k_1 + k_2$ being odd or at $n_1 + n_2$ and $k_1 + k_2$ being even, the integral $I$ is strictly positive, i.e. $I > 0$.

## Proof of Theorem 1

Let us prove the first assertion of Theorem 1. To be definite, we suppose that the quantity $n = n_1 + n_2$ is even and the quantity $k = k_1 + k_2$ is odd. Substituting the expression (1.10) into the integral (1.10), we get

$$I = \sqrt{2^{n+k} n_1! n_2! k_1! k_2!} \sum_{s=0}^{\min(n_1, k_1)} \sum_{l=0}^{\min(n_2, k_2)} \frac{1}{(n_1 - s)!(n_2 - l)!} \frac{1}{(k_1 - s)!(k_2 - l)!} \frac{1}{2^{s+l} s! l!} \times$$

$$\times \int_{-\infty}^{+\infty} e^{-x^2} x^{n-(l+s)} dx \int_{-\infty}^{+\infty} e^{-y^2} y^{k-(l+s)} dy, \quad (1.13)$$

Let us denote $\lambda = l + s$ and consider the double integral from the expression (1.13)

$$\int_{-\infty}^{+\infty} e^{-x^2} x^{n-\lambda} dx \int_{-\infty}^{+\infty} e^{-y^2} y^{k-\lambda} dy \qquad (1.14)$$

The quantity $\lambda$ can be even or odd and varies from 0 to $\min(n_1,k_1)+\min(n_2,k_2)$. Consider both cases of the values of $\lambda$: even and odd ones. If $\lambda$ is even, then the quantity $n-\lambda$ is even, and the quantity $k-\lambda$ is odd. Consequently, the second integral in the expression (1.14) over the variable $y$ equals zero. If $\lambda$ is odd, then the quantity $n-\lambda$ is odd, and the quantity $k-\lambda$ is even. Consequently, the first integral in the expression (1.14) over the variable $x$ equals zero. Thus, for any value of $\lambda$, the integral (1.14) will be equal to zero. Similar reasoning is true for odd values of $n$ and even values of $k$. As a result, the first assertion of Theorem 1 is proved.

Let us prove the second assertion of Theorem 1. Let $n,k$ be even, then for the odd values of $\lambda$ both integrals in the expression (1.14) will vanish. That is, in the sums (1.13) there will be no summands, in which $\lambda = l+s$ is odd. For even values of $\lambda$, the values of $n-\lambda, k-\lambda$ will be even. Consequently, the integrals (1.14) will be non-zero and strictly positive. As a result, in the sum (1.13), only summands with even values of $\lambda$ will be nonzero.

If $n,k$ are odd, then by analogy with the previous case, the nonzero summands are summands for which the values of $\lambda$ are odd. Thus, in both cases, the integral (1.12) will be strictly positive for even and odd values of $n,k$, which was to be proved.

*Theorem 2*

Let the numbers $n = n_1+n_2$, $k = k_1+k_2$ from Theorem 1 be even or odd at the same time, then the integral (1.12) takes the values

$$I_{n,k} = \pi\sqrt{n_1!n_2!k_1!k_2!} \underbrace{\sum_{s=0}^{\min(n_1,k_1)} \sum_{l=0}^{\min(n_2,k_2)}}_{s+l,n,k-even/odd} \frac{|k-(l+s)-1|!!}{s!(n_1-s)!(k_1-s)!} \frac{|n-(l+s)-1|!!}{l!(n_2-l)!(k_2-l)!}, \qquad (1.15)$$

*In this case, the summation in the expression (1.15) is performed over all even values of $s+l$ for even values of $n,k$ and over all odd values of $s+l$ for odd values of $n,k$.*

*Proof of Theorem 2*

The integrals (1.14) can be calculated explicitly. We denote $n-\lambda = 2\nu$, $\nu \in \mathbb{N}_0$ and $k-\lambda = 2\mu$, $\mu \in \mathbb{N}_0$, since the integral (1.14) will be non-zero only for even values of $n-\lambda$ and $k-\lambda$.

$$\int_{-\infty}^{+\infty} e^{-x^2} x^{2n} dx = \sqrt{\pi}\,\frac{|2n-1|!!}{2^n}, \ n \in \mathbb{N}_0,$$

$$\int_{-\infty}^{+\infty} e^{-x^2} x^{2\nu} dx \int_{-\infty}^{+\infty} e^{-y^2} y^{2\mu} dy = \frac{2^{l+s}\pi}{\sqrt{2^{n+k}}} |n-(l+s)-1|!!|k-(l+s)-1|!! \qquad (1.16)$$

Using the expression (1.16) for the integral (1.13), we obtain the representation (1.15), which was to be proved.

*Definition 3*

Let $n, k \in \mathbb{N}_0$, define the modified Kronecker symbol $\tilde{\delta}_{n,k}$ as follows

$$\tilde{\delta}_{n,k} = \begin{cases} 1, & \text{if } n \text{ and } k \text{ are even,} \\ 1, & \text{if } n \text{ and } k \text{ are odd,} \\ 0, & \text{if } n \text{ is even and } k \text{ is odd,} \\ 0, & \text{if } n \text{ is odd and } k \text{ is even.} \end{cases} \quad (1.17)$$

*Corollary 2*

For the polynomials $\mathcal{P}_{n,k}(x,y)$ at $(x,y) \in \mathbb{R}^2$, according to Theorem 1 and Theorem 2, the following representation is true:

$$\int_{-\infty}^{+\infty}\int_{-\infty}^{+\infty} \rho_2(x,y) \mathcal{P}_{n_1,k_1}(x,y) \mathcal{P}_{n_2,k_2}(x,y) dx dy = \mathcal{N}^{(2)}_{n_1+n_2, k_1+k_2} \tilde{\delta}_{n_1+n_2, k_1+k_2}, \quad (1.18)$$

$$\mathcal{N}^{(2)}_{n_1+n_2, k_1+k_2} = \pi \sqrt{n_1! n_2! k_1! k_2!} \underbrace{\sum_{s=0}^{\min(n_1,k_1)} \sum_{l=0}^{\min(n_2,k_2)}}_{s+l, n, k-\text{even/odd}} \frac{|k-(l+s)-1|!!}{s!(n_1-s)!(k_1-s)!} \frac{|n-(l+s)-1|!!}{l!(n_2-l)!(k_2-l)!}.$$

*Theorem 3*

Suppose $n, k \in \mathbb{N}_0$ and $\varsigma_1, \varsigma_2$ are constant values then the following expression is true for the Hermitian polynomials

$$\frac{1}{\sqrt{2^{n+k}\pi n! k!}} \int_{-\infty}^{+\infty} e^{-\varsigma^2} H_n(\varsigma+\varsigma_1) H_k(\varsigma+\varsigma_2) d\varsigma = \mathcal{P}_{n,k}(\varsigma_1, \varsigma_2). \quad (1.19)$$

*Proof of Theorem 3*

We use the formula for the Hermitian polynomials

$$H_n(\varsigma+\xi) = \sum_{k=0}^{n} C_n^k H_k(\varsigma) (2\xi)^{n-k}. \quad (1.20)$$

Substituting (1.20) into the left side of the equation (1.19), we obtain

$$\int_{-\infty}^{+\infty} e^{-\varsigma^2} H_n(\varsigma+\varsigma_1) H_k(\varsigma+\varsigma_2) d\varsigma = \sum_{s=0}^{n}\sum_{l=0}^{k} C_k^l C_n^s (2\varsigma_1)^{n-s} (2\varsigma_2)^{k-l} \int_{-\infty}^{+\infty} e^{-\varsigma^2} H_s(\varsigma) H_l(\varsigma) d\varsigma =$$

$$= \sqrt{\pi} \sum_{s=0}^{n}\sum_{l=0}^{k} C_k^l C_n^s (2\varsigma_1)^{n-s} (2\varsigma_2)^{k-l} 2^s s! \delta_{sl} = 2^n \sqrt{\pi} n! \sum_{s=0}^{\min(n,k)} C_k^s \frac{1}{(n-s)!} \varsigma_1^{n-s} (2\varsigma_2)^{k-s} =$$

$$= 2^n 2^k n! k! \sqrt{\pi} \varsigma_1^n \varsigma_2^k \sum_{s=0}^{\min(n,k)} \frac{1}{s!(k-s)!(n-s)!} \frac{1}{(2\varsigma_1\varsigma_2)^s},$$

According to Corollary 1, the obtained expression proves the theorem.

## Corollary 3

It follows from Theorem 3 that at $n = k$, the formula (1.19) goes into the expression

$$\frac{1}{2^n n! \sqrt{\pi}} \int e^{-\varsigma^2} H_n(\varsigma + \varsigma_1) H_n(\varsigma + \varsigma_2) d\varsigma = \mathcal{P}_{n,n}(\varsigma_1, \varsigma_2) = L_n(-2\varsigma_1 \varsigma_2), \qquad (1.21)$$

where $L_n$ is the Laguerra polynomials.

## Proof of Corollary 3

Indeed, on the one hand, for the Laguerre polynomials the representation [28, 29] is valid

$$L_n(x) = \sum_{k=0}^{n} C_n^{n-k} \frac{(-x)^k}{k!} = \sum_{k=0}^{n} \frac{n!(-x)^k}{(n-k)!k!k!} = \sum_{s=0}^{n} \frac{n!(-x)^{n-s}}{s!(n-s)!(n-s)!} = \sum_{s=0}^{n} \frac{C_n^s}{(n-s)!} (-x)^{n-s}. \qquad (1.22)$$

On the other hand, from the formula (1.20) at $n = k$, we obtain

$$\frac{1}{2^n n! \sqrt{\pi}} \int_{-\infty}^{+\infty} e^{-\varsigma^2} H_n(\varsigma + \varsigma_1) H_n(\varsigma + \varsigma_2) d\varsigma = 2^n n! \varsigma_1^n \varsigma_2^n \sum_{s=0}^{n} \frac{1}{s!(n-s)!(n-s)!} \frac{1}{(2\varsigma_1 \varsigma_2)^s} =$$

$$= \sum_{s=0}^{n} \frac{C_n^s}{(n-s)!} (2\varsigma_1 \varsigma_2)^{n-s} = L_n(-2\varsigma_1 \varsigma_2). \qquad (1.23)$$

Comparing the expressions (1.22) and (1.23), we obtain the validity of the expression (1.21).

## Theorem 4

The elements of the universal density matrix $\mathcal{W}$ are of the form

$$w_{n,k}(x, p) = \frac{(-1)^n}{\pi \hbar} e^{-\kappa^2 x^2 - \frac{p^2}{\hbar^2 \kappa^2}} \mathcal{P}_{n,k}\left(-\kappa x - i\frac{p}{\hbar \kappa}, \kappa x - i\frac{p}{\hbar \kappa}\right), \qquad (1.24)$$

where $\kappa = \sqrt{\frac{m\omega}{\hbar}}$.

## Proof of Theorem 4

Using the formula (1.19), we obtain the expression for the function $w_{n,k}(x, p)$. It follows from the definition (1.4) that

$$w_{n,k}(x, p) = \frac{1}{2\pi \hbar} \frac{1}{\sqrt{2^{n+k} n! k!}} \left(\frac{m\omega}{\pi \hbar}\right)^{\frac{1}{2}} \times$$

$$\times \int e^{-\frac{m\omega\left[\left(x+\frac{s}{2}\right)^2 + \left(x-\frac{s}{2}\right)^2\right]}{2\hbar}} e^{-i\frac{ps}{\hbar}} H_n\left(\sqrt{\frac{m\omega}{\hbar}}\left(x - \frac{s}{2}\right)\right) H_k\left(\sqrt{\frac{m\omega}{\hbar}}\left(x + \frac{s}{2}\right)\right) ds =$$

$$= \frac{1}{2\pi \hbar} \frac{1}{\sqrt{2^{n+k} \pi n! k!}} e^{-\kappa^2 x^2} \int e^{-\frac{\xi^2}{4} - i\frac{p\xi}{\kappa \hbar}} H_n\left(\kappa x - \frac{\xi}{2}\right) H_k\left(\kappa x + \frac{\xi}{2}\right) d\xi, \qquad (1.25)$$

where $\kappa = \sqrt{\dfrac{m\omega}{\hbar}}$, $\xi = s\kappa$. We use the expression

$$-\left(\frac{1}{2}\xi + i\frac{p}{\hbar\kappa}\right)^2 - \frac{p^2}{\hbar^2\kappa^2} = -\varsigma^2 - \frac{p^2}{\hbar^2\kappa^2}. \tag{1.26}$$

Substituting (1.26) into the expression (1.25) and using the property of the Hermitian polynomials $H_n(-x) = (-1)^n H_n(x)$ (that is $H_n\left(-\left(\varsigma - \kappa x - i\dfrac{p}{\hbar\kappa}\right)\right) = (-1)^n H_n\left(\varsigma - \kappa x - i\dfrac{p}{\hbar\kappa}\right)$) we obtain:

$$w_{n,k}(x, p) = \frac{(-1)^n}{\pi\hbar\sqrt{2^{n+k}\pi n! k!}} e^{-\kappa^2 x^2 - \frac{p^2}{\hbar^2\kappa^2}} \int e^{-\varsigma^2} H_n\left(\varsigma - \kappa x - i\frac{p}{\hbar\kappa}\right) H_k\left(\varsigma + \kappa x - i\frac{p}{\hbar\kappa}\right) d\varsigma. \tag{1.27}$$

We denote $\varsigma_1 = \kappa x - i\dfrac{p}{\hbar\kappa}$, $\varsigma_2 = -\kappa x - i\dfrac{p}{\hbar\kappa}$ and use the formula (1.19) for the expression (1.27), we obtain

$$w_{n,k}(x, p) = \frac{(-1)^n}{\pi\hbar} e^{-\kappa^2 x^2 - \frac{p^2}{\hbar^2\kappa^2}} \mathcal{P}_{n,k}\left(-\kappa x - i\frac{p}{\hbar\kappa}, \kappa x - i\frac{p}{\hbar\kappa}\right),$$

which was to be proved.

*Corollary 4*

*It follows from Theorem 4 and Property 1 that the following relations are valid for the polynomials $\mathcal{P}_{n,k}$:*

$$\frac{(-1)^n}{\pi\hbar} \int_{-\infty}^{+\infty} e^{-\kappa^2 x^2 - \frac{p^2}{\hbar^2\kappa^2}} \mathcal{P}_{n,k}\left(-\kappa x - i\frac{p}{\hbar\kappa}, \kappa x - i\frac{p}{\hbar\kappa}\right) dp = \overline{\Psi}_n(x)\Psi_k(x),$$

$$\frac{(-1)^n}{\pi\hbar} \int_{-\infty}^{+\infty} e^{-\kappa^2 x^2 - \frac{p^2}{\hbar^2\kappa^2}} \mathcal{P}_{n,k}\left(-\kappa x - i\frac{p}{\hbar\kappa}, \kappa x - i\frac{p}{\hbar\kappa}\right) dx = \overline{\tilde{\Psi}}_n(p)\tilde{\Psi}_k(p).$$

**Remark**

Note that according to Corollary 3, the expression (1.24) goes into the Wigner functions at $n = k$, which is consistent with Property 2.

From the Hermitian character of the density matrix $\mathcal{W}$ (Property 3) and the expression (1.9), it follows that

$$\overline{W}^T = \left(\overline{C}^T \overline{\mathcal{W}} \overline{C}\right)^T = \left(\overline{\mathcal{W}}\overline{C}\right)^T \overline{C} = \overline{C}^T \left(\overline{\mathcal{W}}\right)^T \overline{C} = \overline{C}^T \mathcal{W}^\dagger \overline{C} = \overline{C}^T \mathcal{W} C = W,$$

and obtained elements $w_{n,k}(x, p)$ (1.24) satisfy Property 3. Indeed, denote $z = \kappa x + i\dfrac{p}{\hbar\kappa}$, then the expression (1.24) takes the form

$$w_{n,k}(x,p) = \frac{(-1)^n}{\pi\hbar} e^{-\kappa^2 x^2 - \frac{p^2}{\hbar^2\kappa^2}} \mathcal{P}_{n,k}(-z,\bar{z}),$$

$$\frac{\mathcal{P}_{n,k}(-z,\bar{z})}{\sqrt{2^{n+k}n!k!}} = \sum_{s=0}^{\min(n,k)} \frac{(-z)^{n-s}\bar{z}^{k-s}}{2^s s!(k-s)!(n-s)!} = (-1)^n z^n \bar{z}^k \sum_{s=0}^{\min(n,k)} \frac{(-1)^s}{2^s s!(k-s)!(n-s)!|z|^{2s}},$$

$$w_{n,k}(x,p) = \frac{\sqrt{2^{n+k}n!k!}}{\pi\hbar} z^n \bar{z}^k e^{-|z|^2} \sum_{s=0}^{\min(n,k)} \frac{(-1)^s}{2^s s!(k-s)!(n-s)!|z|^{2s}}. \qquad (1.28)$$

Let us check the Hermiticity requirement of (1.7), from the representation (1.28) we obtain

$$w_{k,n}(x,p) = \frac{\sqrt{2^{n+k}n!k!}}{\pi\hbar} z^k \bar{z}^n e^{-|z|^2} \sum_{s=0}^{\min(n,k)} \frac{(-1)^s}{2^s s!(k-s)!(n-s)!|z|^{2s}},$$

$$\bar{w}_{k,n}(x,p) = \frac{\sqrt{2^{n+k}n!k!}}{\pi\hbar} \bar{z}^k z^n e^{-|z|^2} \sum_{s=0}^{\min(n,k)} \frac{(-1)^s}{2^s s!(k-s)!(n-s)!|z|^{2s}} = w_{n,k}(x,p).$$

Note that the obtained expressions $w_{n,k}(x,p)$ are complex in the general case (1.28). Only at $n=k$ the diagonal elements $w_{n,n}(x,p)$ are valid. At $n \neq k$ the off-diagonal elements are representable in the form

$$w_{n,k}(x,p) = \frac{\sqrt{2^{n+k}n!k!}}{\pi\hbar} |z|^{2\min(n,k)} e^{-|z|^2} \sum_{s=0}^{\min(n,k)} \frac{(-1)^s}{2^s s!(k-s)!(n-s)!|z|^{2s}} \begin{cases} \bar{z}^{k-\min(n,k)}, k > n, \\ z^{n-\min(n,k)}, k < n, \end{cases} \qquad (1.29)$$

or

$$w_{n,k}(x,p) = \frac{\sqrt{2^{n+k}n!k!}}{\pi\hbar} |z|^{\min(n,k)} e^{-|z|^2} \sum_{s=0}^{\min(n,k)} \frac{(-1)^s}{2^s s!(k-s)!(n-s)!|z|^{2s}} \begin{cases} |z|^k e^{-i[k-\min(n,k)]\varphi}, k > n, \\ |z|^n e^{i[n-\min(n,k)]\varphi}, k < n, \end{cases}$$

or

$$w_{n,k}(x,p) = \frac{\sqrt{2^{n+k}n!k!}}{\pi\hbar} |z|^{n+k} e^{-|z|^2} \sum_{s=0}^{\min(n,k)} \frac{(-1)^s}{2^s s!(k-s)!(n-s)!|z|^{2s}} e^{i(n-k)\varphi},$$

where it is taken into account that $\min(n,k) + \max(n,k) = n+k$. The quantity $|z|^2$ corresponds to the energy (1.6), i.e.

$$|z|^2 = \frac{2}{\hbar\omega}\left(\frac{p^2}{2m} + \frac{m\omega^2 x^2}{2}\right) = 2\varepsilon(x,p). \qquad (1.30)$$

The phase $\varphi$ corresponds to the vectorial angle on the plane of the phase space a $(x,p)$, as

$$\varphi = \arg z = \operatorname{arctg}\left(\frac{p}{m\omega x}\right). \qquad (1.31)$$

The expression (1.31) indicates the important physical significance of the off-diagonal elements of the universal density matrix $\mathcal{W}$ in the phase space. For the quantum harmonic oscillator, $n = k$ the off-diagonal elements are absent, which leads to the constancy of the probability density function on the phase trajectories (1.29), since the influence of the phase $\varphi$ (1.31) in the expression (1.29) is absent

$$W = \begin{pmatrix} w_{1,1} & 0... & 0 \\ 0 & w_{2,2}.. & 0 \\ ... & 0 & ... \end{pmatrix}, \quad \varepsilon = \frac{|z|^2}{2} = const. \tag{1.32}$$

For an arbitrary quantum system in the general case $n \neq k$, therefore, the value of the phase $\varphi$ (1.31) contributes to the probability density function $w_{n,k}(x,p)$ (1.29) along the phase trajectory $\varepsilon = const$. Therefore, the probability density (Wigner function) $W$ (1.9) will be variable along the phase trajectory $\varepsilon = const$.

As can be seen from the expression (1.29), the change in the probability density along the phase trajectory $\varepsilon = const$ will be periodic. The quantity $\varpi_{n,k} = n - k \in \mathbb{Z}$ takes integer values. It follows from the expression (1.29) that the functions $w_{n,k}(x,p)$ will be periodic with a period

$$T_{n,k} = \frac{2\pi}{|\varpi_{n,k}|} = \frac{2\pi}{|n-k|}. \tag{1.33}$$

It follows from (1.33) that the farther from the diagonal of the universal density matrix $\mathcal{W}$, the period is shorter, i.e. the oscillation frequency $|\varpi_{n,k}|$ of the complex probability density function $w_{n,k}(x,p)$ is higher. The closer to the diagonal of the universal density matrix $\mathcal{W}$, the longer the period (1.33) of oscillations of the probability density $w_{n,k}(x,p)$ (the lower the frequency $|\varpi_{n,k}|$). On the diagonal of the universal density matrix $\mathcal{W}$, the oscillation frequency is zero $|\varpi_{n,n}| = 0$ and the probability density becomes real and constant along the phase trajectory $\varepsilon = const$.

When considering the modulus $|w_{n,k}(x,p)|$ of probability density functions, the oscillations disappear, since $|e^{i\varpi_{n,k}\varphi}| = 1$

$$|w_{n,k}(x,p)| = \frac{\sqrt{2^{n+k} n! k!}}{\pi \hbar} |z|^{n+k} e^{-|z|^2} \left| \sum_{s=0}^{\min(n,k)} \frac{(-1)^s}{2^s s!(k-s)!(n-s)!|z|^{2s}} \right|, \tag{1.34}$$

i.e. according to (1.30), on the phase trajectories $\varepsilon = const$

$$|w_{n,k}(x,p)| = const.$$

From the point of view of the theory of complex-variable functions, the elements $w_{n,k}(x,p)$ of the universal density matrix $\mathcal{W}$ are multivalent complex functions on the Riemann surface. The diagonal elements $w_{n,n}(x,p)$ have constant real values on the phase trajectories $\varepsilon = const$ at the phase angles $0 \leq \varphi \leq 2\pi$.

## Definition 4

Let $n, k \in \mathbb{N}_0$ и $x \in \mathbb{R}$. *Define the polynomials* $\Upsilon_{n,k}(x)$ *of power* $n+k$ *as*

$$\Upsilon_{n,k}(x) \stackrel{det}{=} x^{n+k} \sqrt{2^{n+k} n! k!} \sum_{s=0}^{\min(n,k)} \frac{(-1)^s}{2^s s!(k-s)!(n-s)! x^{2s}}. \quad (1.35)$$

## Property 4

*The polynomials* $\Upsilon_{n,k}(x)$ *satisfy the condition*

$$\mathcal{P}_{n,k}(-z, \bar{z}) = (-1)^n \Upsilon_{n,k}(|z|) e^{i(n-k)\varphi}, \quad (1.36)$$

*where* $\varphi = \arg z$ *and the orthogonality condition is fulfilled for them*

$$\int_{-\infty}^{+\infty} \rho_1(x) \Upsilon_{n_1,k_1}(x) \Upsilon_{n_2,k_2}(x) dx = \mathcal{N}^{(1)}_{n_1+n_2,k_1+k_2} \tilde{\delta}_{n_1+n_2,k_1+k_2}, \quad (1.37)$$

$$\mathcal{N}^{(1)}_{n_1+n_2,k_1+k_2} = \sqrt{\pi n_1! n_2! k_1! k_2!} \sum_{s=0}^{\min(n_1,k_1)} \sum_{l=0}^{\min(n_2,k_2)} \frac{(-1)^{s+l} |n_1+n_2+k_1+k_2-2(l+s)-1|!!}{s! l! (k_1-s)!(n_1-s)!(k_2-l)!(n_2-l)!},$$

*where the weight function* $\rho_1(x) = e^{-x^2}$.

## Proof of Property 4

The expression (1.36) is obtained by comparing the expressions (1.35) and (1.28). We calculate the integral (1.37).

$$\int_{-\infty}^{+\infty} \rho_1(x) \Upsilon_{n_1,k_1}(x) \Upsilon_{n_2,k_2}(x) dx =$$

$$= \sum_{s=0}^{\min(n_1,k_1)} \sum_{l=0}^{\min(n_2,k_2)} \frac{(-1)^{s+l}}{2^{s+l} s!(k_1-s)!(n_1-s)! l!(k_2-l)!(n_2-l)!} \sqrt{2^{n+k} n_1! n_2! k_1! k_2!} \int_{-\infty}^{+\infty} e^{-x^2} x^{n_1+n_2+k_1+k_2-2(s+l)} dx. \quad (1.38)$$

We calculate the integral standing in the expression (1.38) using (1.16) and Theorem 2. Denote $n = n_1+n_2$, $k = k_1+k_2$, $\lambda = l+s$. The value $2\lambda$ is even. If the values $n$ and $k$ are even, then the expression $n+k-2\lambda$ is also even. If $n$ and $k$ are odd, then the expression $n+k$ is even and the expression is $n+k-2\lambda$ also even. As a result, at $n$ and $k$ being even and at $n$ and $k$ being odd, the integral $\int_{-\infty}^{+\infty} e^{-x^2} x^{n+k-2\lambda} dx$ will be different from zero.

$$\int_{-\infty}^{+\infty} e^{-x^2} x^{n+k-2\lambda} dx = 2^\lambda \sqrt{\frac{\pi}{2^{n+k}}} |n+k-2\lambda-1|!! \quad (1.39)$$

If the value $n$ or $k$ is odd, then the expression $n+k-2\lambda$ will be odd and the integral $\int_{-\infty}^{+\infty} e^{-x^2} x^{n+k-2\lambda} dx$ will be zero. Substituting (1.39) into (1.38), we obtain

$$\int\limits_{-\infty}^{+\infty} \rho_1(x)\Upsilon_{n_1,k_1}(x)\Upsilon_{n_2,k_2}(x)dx =$$

$$= \sqrt{\pi n_1!n_2!k_1!k_2!} \sum_{s=0}^{\min(n_1,k_1)} \sum_{l=0}^{\min(n_2,k_2)} \frac{(-1)^{s+l}|n+k-2(l+s)-1|!!}{s!l!(k_1-s)!(n_1-s)!(k_2-l)!(n_2-l)!},$$

which was to be proved.

Fig. 1 shows the graphs of the functions $e^{-x^2}\Upsilon_{n,k}(x)$. Fig. 1 illustrates that the polynomials $\Upsilon_{n,k}(x)$ have zeros and are alternating in sign. Regarding the origin of coordinates, the polynomials $\Upsilon_{n,k}(x)$ are even and odd.

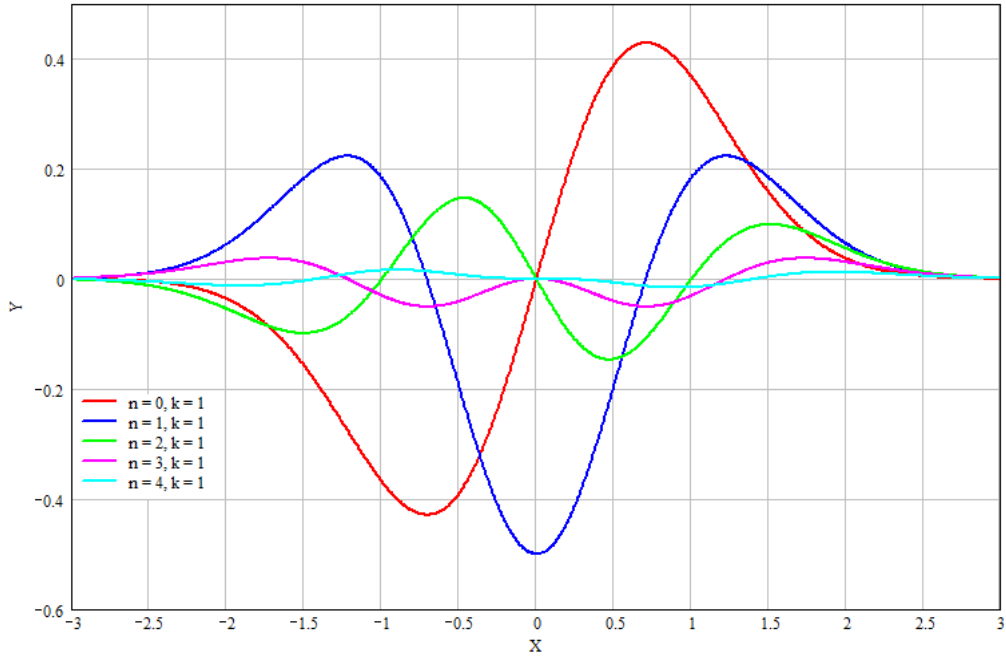

Fig. 1 Graph of the functions $e^{-x^2}\Upsilon_{n,k}(x)/\sqrt{2^{n+k}n!k!}$

We write the expressions for the elements $w_{n,k}(x,p)$, according to (1.29)-(1.31) and (1.35) take the form:

$$w_{n,k}(x,p) = \frac{1}{\pi\hbar}e^{-|z|^2}\Upsilon_{n,k}(|z|)e^{i(n-k)\varphi}, \qquad (1.40)$$

or

$$w_{n,k}(x,p) = \frac{1}{\pi\hbar}e^{-2\varepsilon(x,p)}\Upsilon_{n,k}\left(\sqrt{2\varepsilon(x,p)}\right)e^{i(n-k)\operatorname{arctg}\left(\frac{p}{m\omega x}\right)}.$$

It follows from the expressions (1.40) that when the phase angle $0 \leq \varphi \leq 2\pi$ changes, the values $w_{n,n+1}(x,p)$ are located on one sheet of the Riemann surface. The values of the elements $w_{n,n+2}(x,p)$ cover two sheets of the Riemann surface when the phase angle $0 \leq \varphi \leq 2\pi$ changes. The value of the elements of each next diagonal (upper from the main diagonal) of the universal density matrix $\mathcal{W}$ cover the number of sheets of the Riemann surface equal to the number of the diagonal. A similar process occurs with the elements of the lower diagonals, only the covering goes in the opposite direction.

In [17], [35], when considering the complex principle of least action, we showed that, for microsystems, the phase of the wave function (action) corresponds to mappings of one-sheeted Riemann surfaces. For macrosystems, the range of phase change is larger than $2\pi$ and a transition to multivalent Riemann surfaces occurs.

In the present work, the macrosystem is represented as a set of microsystems (oscillators — the simplest quantum systems), for which a single real axis is sufficient. When considering complex quantum systems (macrosystems), it is necessary to use a multivalent Riemann surface, which is associated with the presence of off-diagonal elements in the universal density matrix $\mathcal{W}$. The presence of off-diagonal elements leads to oscillations of the complex values of the probability density $w_{n,k}(x,p)$ (1.40). When moving along a phase trajectory $\varepsilon = const$, the oscillations are the rotation of «vectors» $w_{n,k}(x,p)$ (of a constant length (1.34)) along the Riemann surface.

## Theorem 5

Let $\rho_{k,n} = c_k \bar{c}_n$ be the matrix elements of the density matrix (1.8) and $\Omega^{(n,k)}(\varphi)$ be the rotation matrix

$$\Omega^{(n,k)}(\varphi) = \begin{pmatrix} \cos(\varpi_{n,k}\varphi) & \sin(\varpi_{n,k}\varphi) \\ -\sin(\varpi_{n,k}\varphi) & \cos(\varpi_{n,k}\varphi) \end{pmatrix}, \quad \varpi_{n,k} = n - k. \qquad (1.41)$$

We define the phase vector for the coefficients $c_k$ (at $|c_k| \neq 0$)

$$\vec{n}_k \stackrel{det}{=} \begin{pmatrix} \cos \alpha_k \\ \sin \alpha_k \end{pmatrix} = \frac{1}{|c_k|} \begin{pmatrix} \operatorname{Re} c_k \\ \operatorname{Im} c_k \end{pmatrix}, \qquad (1.42)$$

$$\alpha_k = \arg c_k,$$

then the Wigner function of the quantum system can be represented in the form:

$$W(x,p) = \frac{e^{-2\varepsilon(x,p)}}{\pi\hbar} \sum_{n,k=0}^{+\infty} |\rho_{k,n}| \Upsilon_{n,k}\left(\sqrt{2\varepsilon(x,p)}\right) \vec{n}_k^T \Omega^{(n,k)}(\varphi) \vec{n}_n, \qquad (1.43)$$

where the phase $\varphi = \varphi(x,p)$ is of the form of (1.31) and corresponds to the vectorial angle of the point $(x,p)$ on the phase plane.

## Proof of Theorem 5

Transform the expression (1.9) for the Wigner function:

$$W = \bar{C}^T \mathcal{W} C = \sum_{n,k=0}^{+\infty} \left[ c_n^{(R)} w_{n,k}^{(R)} c_k^{(R)} - c_n^{(R)} w_{n,k}^{(I)} c_k^{(I)} + c_n^{(I)} w_{n,k}^{(R)} c_k^{(I)} + c_n^{(I)} w_{n,k}^{(I)} c_k^{(R)} \right] +$$

$$+ i \sum_{n,k=0}^{+\infty} \left[ c_n^{(R)} w_{n,k}^{(R)} c_k^{(I)} + c_n^{(R)} w_{n,k}^{(I)} c_k^{(R)} - c_n^{(I)} w_{n,k}^{(R)} c_k^{(R)} + c_n^{(I)} w_{n,k}^{(I)} c_k^{(I)} \right], \qquad (1.44)$$

where the upper superscript «R» means a real part and the superscript «I» means an imaginary part of the element. As far the universal density matrix is an Hermitian one (Property 3) then the relations $w_{n,k} = \bar{w}_{k,n}$, $w_{n,k}^{(R)} + i w_{n,k}^{(I)} = w_{k,n}^{(R)} - i w_{k,n}^{(I)}$ are valid, consequently

$$w_{n,k}^{(R)} = w_{k,n}^{(R)}, \quad w_{n,k}^{(I)} = -w_{k,n}^{(I)},$$

$$W = \sum_{n,k=0}^{+\infty} \left[ w_{n,k}^{(R)} \left( c_n^{(R)} c_k^{(R)} + c_n^{(I)} c_k^{(I)} \right) + 2 c_n^{(I)} w_{n,k}^{(I)} c_k^{(R)} \right] + i \sum_{n,k=0}^{+\infty} w_{n,k}^{(I)} \left( c_n^{(R)} c_k^{(R)} + c_n^{(I)} c_k^{(I)} \right). \quad (1.45)$$

When summing the imaginary part of the expression (1.45), the result will equal zero, since $w_{n,n}^{(I)} = 0$, and the summands $w_{n,k}^{(I)} \left( c_n^{(R)} c_k^{(R)} + c_n^{(I)} c_k^{(I)} \right)$ at $n \neq k$ will be compensated by the summands $w_{k,n}^{(I)} \left( c_k^{(R)} c_n^{(R)} + c_k^{(I)} c_n^{(I)} \right) = -w_{n,k}^{(I)} \left( c_k^{(R)} c_n^{(R)} + c_k^{(I)} c_n^{(I)} \right)$. As a result, the expression (1.45) will have only the real part. In accordance with (1.40) for the function $W$, we obtain

$$W = \frac{e^{-|z|^2}}{\pi \hbar} \sum_{n,k=0}^{+\infty} \Upsilon_{n,k}(|z|) |c_n| |c_k| \times$$
$$\times \left[ \cos(\varpi_{n,k} \varphi)(\cos \alpha_n \cos \alpha_k + \sin \alpha_n \sin \alpha_k) + 2 \sin(\varpi_{n,k} \varphi) \sin \alpha_n \cos \alpha_k \right] = \quad (1.46)$$
$$= \frac{e^{-|z|^2}}{\pi \hbar} \sum_{n,k=0}^{+\infty} \Upsilon_{n,k}(|z|) |c_n| |c_k| \cos \alpha_k \left[ \cos(\varpi_{n,k} \varphi) \cos \alpha_n + \sin(\varpi_{n,k} \varphi) \sin \alpha_n \right] +$$
$$+ \frac{e^{-|z|^2}}{\pi \hbar} \sum_{n,k=0}^{+\infty} \Upsilon_{n,k}(|z|) |c_n| |c_k| \sin \alpha_n \left[ \cos(\varpi_{n,k} \varphi) \sin \alpha_k + \sin(\varpi_{n,k} \varphi) \cos \alpha_k \right].$$

Let us take into account that $\varpi_{n,k} = n - k = -\varpi_{k,n}$, and the polynomials $\Upsilon_{n,k}$ are symmetric with respect to the superscripts $n$ and $k$ by virtue of the definition (1.35), i.e. $\Upsilon_{n,k} = \Upsilon_{k,n}$. As a result, the expression (1.46) can be rewritten as

$$W = \frac{e^{-|z|^2}}{\pi \hbar} \sum_{n,k=0}^{+\infty} \Upsilon_{n,k}(|z|) |\rho_{k,n}| \cos \alpha_k \left[ \cos(\varpi_{n,k} \varphi) \cos \alpha_n + \sin(\varpi_{n,k} \varphi) \sin \alpha_n \right] +$$
$$+ \frac{e^{-|z|^2}}{\pi \hbar} \sum_{n,k=0}^{+\infty} \Upsilon_{n,k}(|z|) |\rho_{k,n}| \sin \alpha_k \left[ \cos(\varpi_{n,k} \varphi) \sin \alpha_n - \sin(\varpi_{n,k} \varphi) \cos \alpha_n \right], \quad (1.47)$$

where it is taken into account that $|\rho_{k,n}|^2 = \rho_{k,n} \bar{\rho}_{k,n} = |c_n|^2 |c_k|^2$. Using the definition of the rotation matrix $\Omega^{(n,k)}(\varphi)$ and the phase vector $\vec{n}_k$, we obtain

$$\vec{n}_k^T \Omega^{(n,k)}(\varphi) \vec{n}_n = (\cos \alpha_k \quad \sin \alpha_k) \begin{pmatrix} \cos(\varpi_{n,k} \varphi) & \sin(\varpi_{n,k} \varphi) \\ -\sin(\varpi_{n,k} \varphi) & \cos(\varpi_{n,k} \varphi) \end{pmatrix} \begin{pmatrix} \cos \alpha_n \\ \sin \alpha_n \end{pmatrix},$$

$$\vec{n}_k^T \Omega^{(n,k)}(\varphi) \vec{n}_n = \cos \alpha_k \left[ \cos(\varpi_{n,k} \varphi) \cos \alpha_n + \sin(\varpi_{n,k} \varphi) \sin \alpha_n \right] + \quad (1.48)$$
$$+ \sin \alpha_k \left[ \cos(\varpi_{n,k} \varphi) \sin \alpha_n - \sin(\varpi_{n,k} \varphi) \cos \alpha_n \right].$$

Substituting (1.48) into the expression (1.47), we obtain the expression (1.43). Theorem 5 is proved.

**Remark**

The phase trajectories (characteristics) $\mathcal{E}(x,p) = \dfrac{p^2}{2m} + U(x) = const$ will correspond to the quantum system, and in the general case these trajectories differ from the phase trajectories of the harmonic oscillator $\varepsilon(x,p) = \dfrac{p^2}{2m} + \dfrac{m\omega^2 x^2}{2} = const$ (1.6). When moving along the phase trajectory $\varepsilon = const$, the rotation matrix $\Omega^{(n,k)}(\varphi)$ will be a unity matrix and the expression (1.43) will go into the Wigner function $W$ (1.6) for the harmonic oscillator. In this case, the probability density $W$ will be constant along the phase trajectories $\varepsilon = const$.

Moving along the phase trajectory $\mathcal{E} = const$, the Wigner function (1.43) will not be constant in the general case. If the phase trajectory $\mathcal{E} = const$ is closed, then due to the periodicity of the rotation matrix $\Omega^{(n,k)}(\varphi)$, the function $W$ (1.43) will also be periodic. The frequency of oscillations of the function $W$ will depend on the frequency of intersection of the phase trajectory $\mathcal{E} = const$ with the same phase trajectory of the harmonic oscillator $\varepsilon = const$.

For the harmonic oscillator, the trajectories $\varepsilon = const$ and $W = const$ coincide, and for an arbitrary quantum system it follows from the expression (1.43) that the trajectories $\mathcal{E} = const$ and $W = const$ are different in the general case.

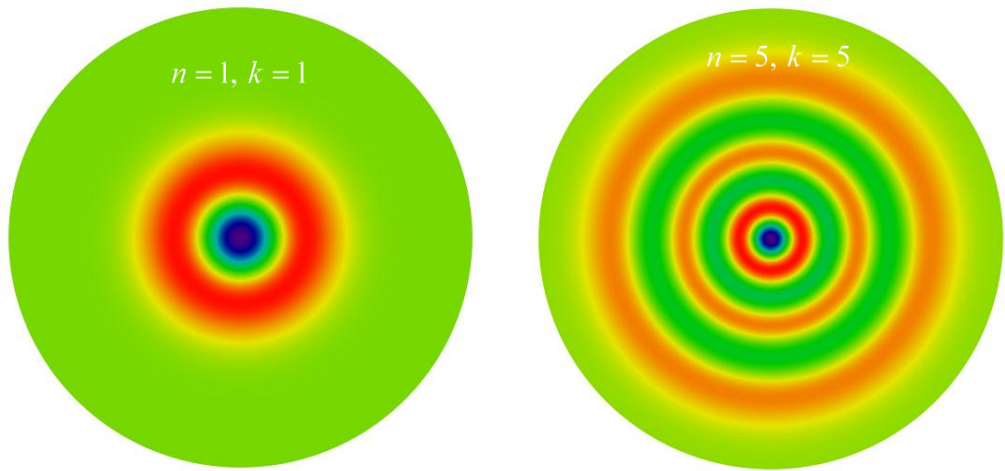

Fig. 2 «Basis» probability density functions at $\varpi_{n,n} = 0$.

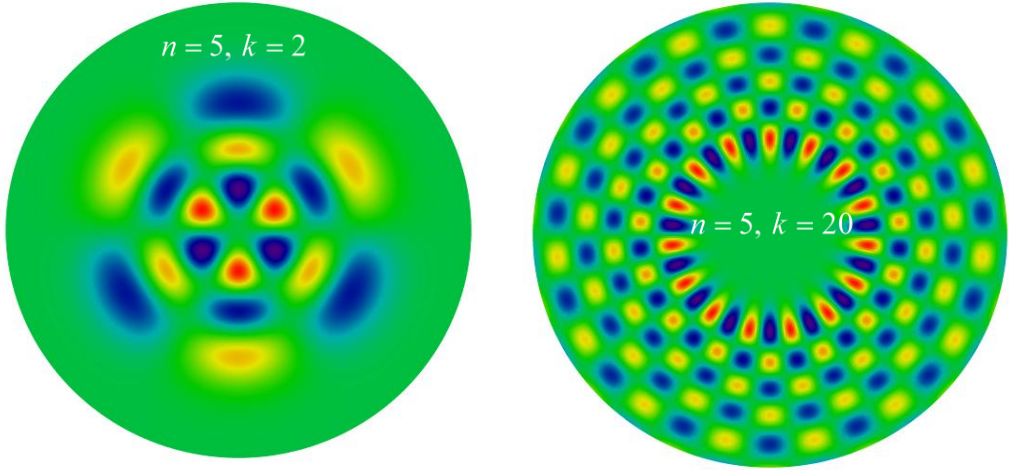

Fig. 3 «Basis» probability density functions at $\varpi_{n,k} \neq 0$.

The expression (1.43) for the Wigner function can be interpreted as an expansion over «basis» functions of the form

$$\mathrm{wc}_{n,k}(x,p) = e^{-2\varepsilon} \Upsilon_{n,k}\left(\sqrt{2\varepsilon}\right) \cos\left(\varpi_{n,k}\varphi\right), \ \mathrm{ws}_{n,k}(x,p) = e^{-2\varepsilon} \Upsilon_{n,k}\left(\sqrt{2\varepsilon}\right) \sin\left(\varpi_{n,k}\varphi\right). \quad (1.49)$$

Figs. 2, 3 show the graphs of the functions $\mathrm{wc}_{n,k}(x,p)$ for different values of $n$ and $k$. In Fig. 2, the functions $\mathrm{wc}_{1,1}$ and $\mathrm{wc}_{5,5}$ correspond to the main diagonal of the universal density matrix $\mathcal{W}$ and have the frequency $\varpi_{n,n} = 0$. The functions $\mathrm{wc}_{n,n}$ and $\mathrm{ws}_{n,n}$ are constant on the phase trajectories $\varepsilon = const$ (coaxial circles), which is observed in Fig. 2. The diagonal basic functions (1.49) make a significant contribution to the sum (1.43) for systems similar to the harmonic oscillator.

In Fig. 3 the functions $\mathrm{wc}_{n,k}$ have non-zero frequencies $\varpi_{n,k} \neq 0$ ($\varpi_{5,2} = 3$ and $\varpi_{5,20} = -15$), which leads to oscillations along the phase trajectories $\varepsilon = const$. The functions $\mathrm{wc}_{n,k}$ with frequencies $\varpi_{n,k} \neq 0$ correspond to the upper and lower diagonals of the universal density matrix $\mathcal{W}$. The further the diagonal from the central diagonal, the higher the frequency $\varpi_{n,k}$ of oscillations. The off-diagonal basic functions (1.49) contribute significantly for systems that are distinctly different from the harmonic oscillator.

**§2 Moyal and Vlasov equations**

From Theorem 5 one can see that the probability density $W$ in the general case will be variable on the phase trajectory $\mathcal{E} = const$. The Wigner function $W$ satisfies the Moyal equation [26]

$$\frac{\partial W}{\partial t} + \frac{1}{m}(\vec{p}, \nabla_r)W - (\nabla_r U, \nabla_p W) = \sum_{l=1}^{+\infty} \frac{(-1)^l (\hbar/2)^{2l}}{(2l+1)!} U\left(\overleftarrow{\nabla}_r, \overrightarrow{\nabla}_p\right)^{2l+1} W, \quad (2.1)$$

where $U$ is potential energy. The Moyal equation (2.1) is a special case of the second Vlasov equation (2.2) for the probability density funciton $f_2(\vec{r},\vec{v},t)$ [20, 21]:

$$\frac{\partial f_2}{\partial t} + \text{div}_r \left[ \vec{v} f_2 \right] + \text{div}_v \left[ \langle \dot{\vec{v}} \rangle f_2 \right] = 0, \qquad (2.2)$$

where

$$\langle \dot{\vec{v}} \rangle (\vec{r}, \vec{v}, t) = \frac{\int\limits_{(\infty)} \dot{\vec{v}} f_3 (\vec{r}, \vec{v}, \dot{\vec{v}}, t) d^3 \dot{v}}{f_2 (\vec{r}, \vec{v}, t)}, \qquad (2.3)$$

$$f_1(\vec{r},t) = \int\limits_{(\infty)} f_2(\vec{r},\vec{v},t) d^3 v = \int\limits_{(\infty)} \int\limits_{(\infty)} f_3(\vec{r},\vec{v},\dot{\vec{v}},t) d^3 \dot{v} d^3 v,$$

$$\frac{\partial f_1}{\partial t} + \text{div}_r \left[ \langle \vec{v} \rangle f_1 \right] = 0, \qquad \langle \vec{v} \rangle (\vec{r},t) = \frac{\int\limits_{(\infty)} \vec{v} f_2 (\vec{r}, \vec{v}, t) d^3 v}{f_1 (\vec{r},t)}. \qquad (2.4)$$

The second Vlasov equation (2.2), as well as the first equation (2.4), belongs to the infinite self-linking chain of the Vlasov equations for the probability density functions $f_1(\vec{r},t)$, $f_2(\vec{r},\vec{v},t)$, $f_3(\vec{r},\vec{v},\dot{\vec{v}},t),\ldots$ The equations in the Vlasov chain are kinematic and are obtained from first principles. To solve the equation (2.2), it is necessary to break the chain of the Vlasov equations and approximate the kinematic quantity $\langle \dot{\vec{v}} \rangle$ (2.3) by the dynamic characteristic.

***Definition 5***

*Dynamic approximation of the mean acceleration flow $\langle \dot{\vec{v}} \rangle$ (2.3) of the form*

$$\langle \dot{v}_\alpha \rangle = \sum_{n=0}^{+\infty} \frac{(-1)^{n+1} (\hbar/2)^{2n}}{m^{2n+1} (2n+1)!} \frac{\partial^{2n+1} U}{\partial x_\alpha^{2n+1}} \frac{1}{f_2} \frac{\partial^{2n} f_2}{\partial v_\alpha^{2n}}, \qquad (2.5)$$

*where superscript $\alpha = 1, 2, 3$ denotes components of the vectors of the coordinate and velocity will be called the Vlasov-Moyal approximation.*

***Property 4***

*In the general case, the vector field of the mean acceleration flow $\langle \dot{\vec{v}} \rangle$ in the Vlasov-Moyal approximation (2.5) has non-zero sources of dissipation $Q_2 = \text{div}_v \langle \dot{\vec{v}} \rangle$*

$$Q_2 = \frac{\partial}{\partial v_\beta} \langle \dot{v}_\beta \rangle = \sum_{n=1}^{+\infty} \frac{(-1)^{n+1} (\hbar/2)^{2n}}{m^{2n+1} (2n+1)!} \frac{\partial^{2n+1} U}{\partial x_\beta^{2n+1}} \frac{\partial}{\partial v_\beta} \left( \frac{1}{f_2} \frac{\partial^{2n} f_2}{\partial v_\beta^{2n}} \right), \qquad (2.6)$$

*where the summing is performed over the repeated superscript $\beta$.*

***Property 5***

*The mean value of $\langle \langle \dot{\vec{v}} \rangle \rangle (\vec{r},t)$ for the Vlasov-Moyal approximation is of the form*

$$\langle \langle \dot{v}_\alpha \rangle \rangle = -\frac{1}{m} \frac{\partial U}{\partial x_\alpha}. \qquad (2.7)$$

*Proof of Property 5*

The mean value of $\langle\langle\dot{\vec{v}}\rangle\rangle(\vec{r},t)$ is of the form:

$$f_1(\vec{r},t)\langle\langle\dot{\vec{v}}\rangle\rangle(\vec{r},t) = \int_{(\infty)} \langle\dot{\vec{v}}\rangle(\vec{r},\vec{v},t) f_2(\vec{r},\vec{v},t) d^3v. \qquad (2.8)$$

Substituting (2.5) into the expression (2.8), we obtain

$$f_1\langle\langle\dot{v}_\alpha\rangle\rangle = \int_{(\infty)} \langle\dot{v}_\alpha\rangle f_2 d^3v = -\frac{1}{m}\frac{\partial U}{\partial x_\alpha} f_1 + \frac{(\hbar/2)^2}{3!m^3}\frac{\partial^3 U}{\partial x_\alpha^3}\int_{(\infty)}\frac{\partial^2 f_2}{\partial v_\alpha^2}d^3v - \frac{(\hbar/2)^4}{5!m^5}\frac{\partial^5 U}{\partial x_\alpha^5}\int_{(\infty)}\frac{\partial^4 f_2}{\partial v_\alpha^4}d^3v + \ldots$$

$$\ldots + \frac{(-1)^{n+1}(\hbar/2)^{2n}}{m^{2n+1}(2n+1)!}\frac{\partial^{2n+1}U}{\partial x_\alpha^{2n+1}}\int_{(\infty)}\frac{\partial^{2n}f_2}{\partial v_\alpha^{2n}}d^3v + \ldots = -\frac{1}{m}\frac{\partial U}{\partial x_\alpha}f_1,$$

where it is taken into consideration that the probability density function equals zero at infinity. Property 5 is proved.

*Theorem 6*

The Vlasov-Moyal approximation transforms the second Vlasov equation (2.2) into the Moyal equation (2.1).

*Proof of Theorem 6*

Substituting the Vlasov-Moyal approximation (2.5) and the expression (2.6) for the sources $Q_2$ into the Vlasov equation (2.2), we obtain

$$\frac{\partial f_2}{\partial t} + (\vec{v}, \nabla_r f_2) + (\langle\dot{\vec{v}}\rangle, \nabla_v f_2) = -f_2 \operatorname{div}_v \langle\dot{\vec{v}}\rangle = -f_2 Q_2,$$

$$\frac{\partial f_2}{\partial t} + (\vec{v}, \nabla_r f_2) - \frac{1}{m}(\nabla_r U, \nabla_v f_2) =$$

$$= -\sum_{n=1}^{+\infty} \frac{(-1)^{n+1}(\hbar/2)^{2n}}{m^{2n+1}(2n+1)!}\frac{\partial^{2n+1}U}{\partial x_\beta^{2n+1}}\left[f_2\frac{\partial}{\partial v_\beta}\left(\frac{1}{f_2}\frac{\partial^{2n}f_2}{\partial v_\beta^{2n}}\right) + \frac{1}{f_2}\frac{\partial f_2}{\partial v_\beta}\frac{\partial^{2n}f_2}{\partial v_\beta^{2n}}\right], \qquad (2.9)$$

Taking into account that $f_2\dfrac{\partial}{\partial v_\beta}\left(\dfrac{1}{f_2}\dfrac{\partial^{2n}f_2}{\partial v_\beta^{2n}}\right) + \dfrac{1}{f_2}\dfrac{\partial f_2}{\partial v_\beta}\dfrac{\partial^{2n}f_2}{\partial v_\beta^{2n}} = \dfrac{\partial^{2n+1}f_2}{\partial v_\beta^{2n+1}}$, the expression (2.9) takes the form

$$\frac{\partial f_2}{\partial t} + (\vec{v}, \nabla_r f_2) - \frac{1}{m}(\nabla_r U, \nabla_v f_2) = \sum_{n=1}^{+\infty} \frac{(-1)^n (\hbar/2)^{2n}}{m^{2n+1}(2n+1)!}\frac{\partial^{2n+1}U}{\partial x_\beta^{2n+1}}\frac{\partial^{2n+1}f_2}{\partial v_\beta^{2n+1}},$$

which goes into the Moyal equation (2.1) when replacing $\vec{p} = m\vec{v}$ and $W(\vec{r},\vec{p},t) = W(\vec{r},m\vec{v},t) = f_2(\vec{r},\vec{v},t)$. Theorem 6 is proved.

**Remark**

The representation (2.7) was used by A. Vlasov in solving the equation (2.2). In particular, in [20, 21] A. Vlasov used the classical limit ($\hbar \to 0$) of the approximation (2.5) instead of the approximation (2.5):

$$\langle \dot{v}_\alpha \rangle = -\frac{1}{m}\frac{\partial U}{\partial x_\alpha}. \qquad (2.10)$$

Substituting (2.10) into the equation (2.2), A. Vlasov obtained the equation known in plasma physics

$$\frac{\partial f_2}{\partial t} + (\vec{v}, \nabla_r f_2) - \frac{1}{m}(\nabla_r U, \nabla_v f_2) = 0, \qquad (2.11)$$

for which, according to Property 4 (2.6), there are no sources of dissipations $Q_2$, that is $Q_2 = 0$. The equation (2.10) corresponds to the equation of motion [20, 21, 35, 25] with a mean value of $\langle\langle \dot{v}_\alpha \rangle\rangle$ (2.7)

$$\frac{d}{dt}\langle v_\alpha \rangle = \left(\frac{\partial}{\partial t} + \langle v_\beta \rangle \frac{\partial}{\partial x_\beta}\right)\langle v_\alpha \rangle = -\frac{1}{f_1}\frac{\partial P_{\alpha\beta}}{\partial x_\beta} + \langle\langle \dot{v}_\alpha \rangle\rangle, \qquad (2.12)$$

where $P_{\alpha\beta}$ is the surface tension tensor

$$P_{\alpha\beta} = \int_{(\infty)} f_2 (v_\alpha - \langle v_\alpha \rangle)(v_\beta - \langle v_\beta \rangle) d^3v,$$

which transforms into the classical or quantum pressure $Q_1 = -\frac{\hbar^2}{2m}\frac{\Delta_r \sqrt{f_1}}{\sqrt{f_1}}$ [35, 24] in the particular case in the «pilot-wave» theory of d'Broglie-Bohm [36-39]. With the approximation (2.10) (no dissipations $Q_2 = 0$), the solution of the equation (2.11) can be represented in the form of characteristics $\mathcal{E} = const$

$$f_2(\vec{r}, \vec{v}, t) = F_2(\mathcal{E}), \qquad (2.13)$$

where $F_2$ is a certain function, and the energy $\mathcal{E}$ is defined by the expression $\mathcal{E} = \frac{mv^2}{2} + U$ and is constant $\mathcal{E} = const$ when moving along the phase trajectory (characteristic). The probability density $f_2$ is also constant when moving along the phase trajectory. Such conservative systems are often found in classical physics. For quantum systems whose potential $U$ satisfies the condition $\frac{\partial^{2n+1} U}{\partial x_\beta^{2n+1}} = 0$ at $n > 0$, there will also be no dissipations $Q_2 = 0$ and the solution of the equation (2.2) can be found in the form (2.13). An example of such a system is the quantum harmonic oscillator with the potential $U = \frac{m\omega^2 x^2}{2}$ and $\varepsilon = const$ (1.6), considered in §1.

In the general case, when using the Vlasov-Moyal approximation, the Vlasov equation (2.2) is dissipative, since the sources of dissipations (2.6) $Q_2 \neq 0$ and the equation (2.2) is a modified Vlasov equation [25]. In this case, for the equation (2.2), one can obtain the equation for the Boltzmann $H_2$-function [25, 35]

$$\Pi_2 S_2 = -Q_2, \qquad \frac{dH_2}{dt} = \langle\langle Q_2 \rangle\rangle, \qquad (2.14)$$

$$H_2(t) = -\int_{(\infty)}\int_{(\infty)} f_2(\vec{r},\vec{v},t) S_2(\vec{r},\vec{v},t) d^3r d^3v = -\langle\langle S_2 \rangle\rangle, \quad S_2 = \ln f_2,$$

$$\Pi_2 = \frac{\partial}{\partial t} + v_\beta \frac{\partial}{\partial x_\beta} + \langle \dot{v}_\beta \rangle \frac{\partial}{\partial v_\beta}.$$

Due to the presence of dissipation $Q_2 \neq 0$ (2.6), the probability density function $f_2 = W$ is not constant on the phase trajectories $\mathcal{E} = const$. In §1, it was shown that the Wigner function $W$ has oscillations with the frequency $\varpi_{n,k}$ on the phase trajectories $\mathcal{E} = const$. The Boltzmann $H_2$-function will also change for nonzero values of $\langle\langle Q_2 \rangle\rangle$ according to the equation (2.14).

### Theorem 7

In the case of the Vlasov-Moyal approximation, the expressions for the mean values of the sources of dissipations $\langle Q_2 \rangle$ and $\langle\langle Q_2 \rangle\rangle$ are of the form

$$\langle Q_2 \rangle = \sum_{n=1}^{+\infty} \frac{(-1)^n (\hbar/2)^{2n}}{m^{2n+1}(2n+1)!} \frac{\partial^{2n+1} U}{\partial x_\beta^{2n+1}} \left\langle \frac{\partial^{2n+1} S_2}{\partial v_\beta^{2n+1}} \right\rangle, \qquad (2.15)$$

$$\langle\langle Q_2 \rangle\rangle = \sum_{n=1}^{+\infty} \frac{(-1)^n (\hbar/2)^{2n}}{m^{2n+1}(2n+1)!} \left\langle \frac{\partial^{2n+1} U}{\partial x_\beta^{2n+1}} \left\langle \frac{\partial^{2n+1} S_2}{\partial v_\beta^{2n+1}} \right\rangle \right\rangle.$$

### Proof of Theorem 7

Let us calculate the mean value of $\langle Q_2 \rangle$. From the expression (2.6) we obtain

$$f_1 \langle Q_2 \rangle = \int_{(\infty)} f_2 Q_2 d^3v = \sum_{n=1}^{+\infty} \frac{(-1)^{n+1} (\hbar/2)^{2n}}{m^{2n+1}(2n+1)!} \frac{\partial^{2n+1} U}{\partial x_\beta^{2n+1}} \int_{(\infty)} \frac{\partial^{2n+1} f_2}{\partial v_\beta^{2n+1}} d^3v +$$

$$+ \sum_{n=1}^{+\infty} \frac{(-1)^n (\hbar/2)^{2n}}{m^{2n+1}(2n+1)!} \frac{\partial^{2n+1} U}{\partial x_\beta^{2n+1}} \int_{(\infty)} \frac{1}{f_2} \frac{\partial f_2}{\partial v_\beta} \frac{\partial^{2n} f_2}{\partial v_\beta^{2n}} d^3v \qquad (2.16)$$

The first integral in the expression (2.16) equals zero. We calculate the second integral

$$\int_{(\infty)} \frac{1}{f_2} \frac{\partial f_2}{\partial v_\beta} \frac{\partial^{2n} f_2}{\partial v_\beta^{2n}} d^3v = \int_{(\infty)} \frac{\partial S_2}{\partial v_\beta} \frac{\partial^{2n} f_2}{\partial v_\beta^{2n}} d^3v = \left. \frac{\partial S_2}{\partial v_\beta} \frac{\partial^{2n-1} f_2}{\partial v_\beta^{2n-1}} \right|_\infty - \int_{(\infty)} \frac{\partial^2 S_2}{\partial v_\beta^2} \frac{\partial^{2n-1} f_2}{\partial v_\beta^{2n-1}} d^3v =$$

$$= -\int_{(\infty)} \frac{\partial^2 S_2}{\partial v_\beta^2} \frac{\partial^{2n-1} f_2}{\partial v_\beta^{2n-1}} d^3v, \qquad (2.17)$$

where it is taken into account that the partial derivatives $\dfrac{\partial^{2n} f_2}{\partial v_\beta^{2n}}$ tend to at infinity zero rather quickly [20, 21]. Repeating the procedure (2.17) $k$ – times ($k = 2n$), we obtain

$$\int_{(\infty)} \frac{1}{f_2} \frac{\partial f_2}{\partial v_\beta} \frac{\partial^{2n} f_2}{\partial v_\beta^{2n}} d^3 v = (-1)^k \int_{(\infty)} \frac{\partial^{k+1} S_2}{\partial v_\beta^{k+1}} \frac{\partial^{2n-k} f_2}{\partial v_\beta^{2n-k}} d^3 v = (-1)^{2n} \int_{(\infty)} f_2 \frac{\partial^{2n+1} S_2}{\partial v_\beta^{2n+1}} d^3 v = f_1 \left\langle \frac{\partial^{2n+1} S_2}{\partial v_\beta^{2n+1}} \right\rangle,$$

$$\langle Q_2 \rangle = \sum_{n=1}^{+\infty} \frac{(-1)^n (\hbar/2)^{2n}}{m^{2n+1} (2n+1)!} \frac{\partial^{2n+1} U}{\partial x_\beta^{2n+1}} \left\langle \frac{\partial^{2n+1} S_2}{\partial v_\beta^{2n+1}} \right\rangle. \tag{2.17}$$

Based on the expression (2.17), the mean value of $\langle\langle Q_2 \rangle\rangle$ will take the form

$$N(t)\langle\langle Q_2 \rangle\rangle(t) = \int_{(\infty)} f_1 \langle Q_2 \rangle d^3 r = \sum_{n=1}^{+\infty} \frac{(-1)^n (\hbar/2)^{2n}}{m^{2n+1} (2n+1)!} \int_{(\infty)} f_1 \frac{\partial^{2n+1} U}{\partial x_\beta^{2n+1}} \left\langle \frac{\partial^{2n+1} S_2}{\partial v_\beta^{2n+1}} \right\rangle d^3 r =$$

$$= N(t) \sum_{n=1}^{+\infty} \frac{(-1)^n (\hbar/2)^{2n}}{m^{2n+1} (2n+1)!} \left\langle \frac{\partial^{2n+1} U}{\partial x_\beta^{2n+1}} \left\langle \frac{\partial^{2n+1} S_2}{\partial v_\beta^{2n+1}} \right\rangle \right\rangle,$$

where $N(t)$ is the number of particles or the normalization coefficient for the probability density function. Theorem 7 is proved.

**§3 Quantum system energy**

We obtain the expression for the energy of an arbitrary quantum system $\langle\langle \mathcal{E} \rangle\rangle$. Let us write the expression for the energy $\mathcal{E}(x,p)$ in the form of the energy of the harmonic oscillator $\varepsilon(x,p)$ and some additional energy:

$$\mathcal{E}(x,p) = \frac{p^2}{2m} + U(x) = \hbar\omega\varepsilon(x,p) + \delta U(x), \tag{3.1}$$

$$\delta U(x) = U(x) - \frac{m\omega^2 x^2}{2}.$$

The function $\delta U(x)$ determines the deviation of the potential energy of an arbitrary system from the potential energy of the harmonic oscillator. The total energy of the system $\langle\langle \mathcal{E} \rangle\rangle$ can be determined by the formula

$$\langle\langle \mathcal{E} \rangle\rangle = \int_{-\infty}^{+\infty} \int_{-\infty}^{+\infty} \mathcal{E}(x,p) W(x,p) dx dp, \tag{3.2}$$

where $W(x,p)$ is the probability density function in the phase space, to which the expression (1.43) corresponds in this case.

***Theorem 8***

Let the energy $\mathcal{E}(x,p)$ of a quantum system be represented in the form (3.1) and the function $\delta U(x)$ allows expansion in a power series with coefficients $a_l$, $l \in \mathbb{N}_0$, then the total energy $\langle\langle \mathcal{E} \rangle\rangle$ of the system (3.2) is of the form:

$$\langle\langle \mathcal{E} \rangle\rangle = \sum_{n=0}^{+\infty} |\rho_{n,n}| \langle\langle \varepsilon_n \rangle\rangle + \qquad (3.3)$$

$$+ \sum_{n,k=0}^{+\infty} |\rho_{k,n}| \cos(\alpha_k - \alpha_n) \sqrt{2^{n+k} n! k!} \sum_{\substack{\frac{n-k+l}{2} \in \mathbb{Z},\, l \geq |n-k|}}^{+\infty} a_l \left(\frac{\hbar}{4m\omega}\right)^{l/2} C_l^{\frac{n-k+l}{2}} \sum_{s=0}^{\min(n,k)} \frac{(-1)^s \left(\frac{k+n+l}{2} - s\right)!}{2^s s! (k-s)! (n-s)!},$$

where $\rho_{k,n} = c_k \bar{c}_n$ is the density matrix elements (1.8); $\alpha_k = \arg c_k$ (1.42); $C_n^k = \dfrac{n!}{k!(n-k)!}$ is the number of combinations; $\langle\langle \varepsilon_n \rangle\rangle = \hbar\omega\left(n + \dfrac{1}{2}\right)$ is the energy levels of the harmonic oscillator.

***Proof of Theorem 8***

According to the expressions (1.30), (1.31), we introduce the notation

$$x = \sqrt{\frac{2\hbar}{m\omega}} x', \quad p = p'\sqrt{2m\hbar\omega}, \quad x'^2 + p'^2 = \varepsilon(x,p), \quad \varphi = \operatorname{arctg}\left(\frac{p'}{x'}\right) = \operatorname{arctg}\left(\frac{p}{m\omega x}\right),$$

$$dxdp = 2\hbar dx'dp' = 2\hbar\sqrt{\varepsilon}d\sqrt{\varepsilon}d\varphi = \hbar d\varepsilon d\varphi. \qquad (3.4)$$

With the use of the notations (3.4), the integral (3.2) takes the form

$$\langle\langle \mathcal{E} \rangle\rangle = \hbar \int_0^{2\pi} d\varphi \int_0^{+\infty} \mathcal{E}W d\varepsilon = \hbar^2 \omega \int_0^{2\pi} d\varphi \int_0^{+\infty} W\varepsilon d\varepsilon + \hbar \int_0^{2\pi} d\varphi \int_0^{+\infty} W\delta U d\varepsilon = I_1 + I_2 \qquad (3.5)$$

Let us consider each integral in the expression (3.5) separately

$$\int_0^{2\pi} d\varphi \int_0^{+\infty} W\varepsilon d\varepsilon = \frac{1}{\pi\hbar} \sum_{n,k=0}^{+\infty} |\rho_{k,n}| \int_0^{2\pi} \vec{n}_k^T \Omega^{(n,k)}(\varphi) \vec{n}_n \, d\varphi \int_0^{+\infty} \Upsilon_{n,k}\left(\sqrt{2\varepsilon}\right) e^{-2\varepsilon} \varepsilon d\varepsilon =$$

$$= \frac{2}{\hbar} \sum_{n,k=0}^{+\infty} |\rho_{k,n}| \delta_{n,k} \vec{n}_k^T \vec{n}_n \int_0^{+\infty} \Upsilon_{n,k}\left(\sqrt{2\varepsilon}\right) e^{-2\varepsilon} \varepsilon d\varepsilon = \frac{2}{\hbar} \sum_{n=0}^{+\infty} |\rho_{n,n}| \int_0^{+\infty} \Upsilon_{n,n}\left(\sqrt{2\varepsilon}\right) e^{-2\varepsilon} \varepsilon d\varepsilon, \qquad (3.6)$$

where it is taken into account that $\int_0^{2\pi} \vec{n}_k^T \Omega^{(n,k)}(\varphi) \vec{n}_n \, d\varphi = 2\pi \delta_{n,k} \vec{n}_k^T \vec{n}_n$. Using the expressions (1.36) and (1.21) from (3.6), we obtain

$$\Upsilon_{n,n}(|z|) = (-1)^n \mathcal{P}_{n,n}(-z,\bar{z}) = (-1)^n L_n\left(2|z|^2\right)$$

$$\int_0^{2\pi} d\varphi \int_0^{+\infty} W\varepsilon d\varepsilon = \frac{2}{\hbar} \sum_{n=0}^{+\infty} (-1)^n |\rho_{n,n}| \int_0^{+\infty} L_n(4\varepsilon) e^{-2\varepsilon} \varepsilon d\varepsilon = \frac{2}{\hbar} \sum_{n=0}^{+\infty} (-1)^n |\rho_{n,n}| (-1)^n \frac{2n+1}{4},$$

$$I_1 = \hbar^2 \omega \int_0^{2\pi} d\varphi \int_0^{+\infty} W\varepsilon \, d\varepsilon = \hbar\omega \sum_{n=0}^{+\infty} |\rho_{n,n}| \left(n + \frac{1}{2}\right) = \sum_{n=0}^{+\infty} |\rho_{n,n}| \langle\langle \varepsilon_n \rangle\rangle. \qquad (3.7)$$

The integral $I_1$ contains the contribution of the diagonal elements of the universal density matrix $\mathcal{W}$. We calculate the second integral $I_2$. Under hypothesis of the theorem, the potential $\delta U(x)$ is expanded into a power series, consequently

$$\delta U(x) = \sum_{l=0}^{+\infty} a_l x^l = \sum_{l=0}^{+\infty} a_l \left(\frac{2\hbar}{m\omega}\right)^{l/2} x'^l = \sum_{l=0}^{+\infty} a_l \left(\frac{2\hbar}{m\omega}\right)^{l/2} \varepsilon^{\frac{l}{2}} \cos^l \varphi, \qquad (3.8)$$

where according to (3.4), $x' = \sqrt{\varepsilon} \cos\varphi$. Substituting the expression (3.8) into the integral $I_2$, we obtain

$$\int_0^{2\pi} d\varphi \int_0^{+\infty} W \delta U \, d\varepsilon =$$
$$= \frac{1}{\pi\hbar} \sum_{n,k=0}^{+\infty} |\rho_{k,n}| \sum_{l=0}^{+\infty} a_l \left(\frac{2\hbar}{m\omega}\right)^{l/2} \int_0^{2\pi} \vec{n}_k^T \Omega^{(n,k)}(\varphi) \vec{n}_n \cos^l \varphi \, d\varphi \int_0^{+\infty} e^{-2\varepsilon} \varepsilon^{\frac{l}{2}} \Upsilon_{n,k}\left(\sqrt{2\varepsilon}\right) d\varepsilon. \qquad (3.9)$$

The integral over an angular variable $\varphi$ in the expression (3.9) can be calculated explicitly

$$\int_0^{2\pi} \Omega^{(n,k)}(\varphi) \cos^l \varphi \, d\varphi = \begin{pmatrix} 1 & 0 \\ 0 & 1 \end{pmatrix} \begin{cases} \dfrac{2\pi}{2^l} C_l^{\frac{\varpi_{n,k}+l}{2}}, & \text{if } \dfrac{\varpi_{n,k}+l}{2} \in \mathbb{Z},\ l \geq |\varpi_{n,k}|, \\ 0, & \text{otherwise.} \end{cases} \qquad (3.10)$$

where the following is taken into account

$$\int_0^{2\pi} \sin(\varpi_{n,k}\varphi)\cos^l\varphi \, d\varphi = 0, \quad \int_0^{2\pi} \cos(\varpi_{n,k}\varphi)\cos^l\varphi \, d\varphi = \begin{cases} \dfrac{2\pi}{2^l} C_l^{\frac{\varpi_{n,k}+l}{2}}, & \text{if } \dfrac{\varpi_{n,k}+l}{2} \in \mathbb{Z},\ l \geq |\varpi_{n,k}|, \\ 0, & \text{otherwise.} \end{cases}$$

As a result, the integral over the variable $\varphi$ is of the form

$$\int_0^{2\pi} \vec{n}_k^T \Omega^{(n,k)}(\varphi) \vec{n}_n \cos^l \varphi \, d\varphi = \cos(\alpha_k - \alpha_n) \begin{cases} \dfrac{2\pi}{2^l} C_l^{\frac{\varpi_{n,k}+l}{2}}, & \text{if } \dfrac{\varpi_{n,k}+l}{2} \in \mathbb{Z},\ l \geq |\varpi_{n,k}|, \\ 0, & \text{otherwise,} \end{cases} \qquad (3.11)$$

where it is taken into account that $\vec{n}_k^T \vec{n}_n = \cos\alpha_k \cos\alpha_n + \sin\alpha_k \sin\alpha_n = \cos(\alpha_k - \alpha_n)$. We consider the integral over a variable $\varepsilon$ in the expression (3.9). In the view of

$$\Upsilon_{n,k}\left(\sqrt{2\varepsilon}\right) = (2\varepsilon)^{\frac{n+k}{2}} \sqrt{2^{n+k} n! k!} \sum_{s=0}^{\min(n,k)} \frac{(-1)^s}{2^s s!(k-s)!(n-s)!(2\varepsilon)^s}. \qquad (3.12),$$

we obtain

$$\int_0^{+\infty} e^{-2\varepsilon} \varepsilon^{\frac{l}{2}} \Upsilon_{n,k}\left(\sqrt{2\varepsilon}\right) d\varepsilon = 2^{n+k} \sqrt{n!k!} \int_0^{+\infty} e^{-2\varepsilon} \varepsilon^{\frac{n+k+l}{2}} \sum_{s=0}^{\min(n,k)} \frac{(-1)^s}{4^s s!(k-s)!(n-s)!\varepsilon^s} d\varepsilon =$$

$$= 2^{n+k} \sqrt{n!k!} \sum_{s=0}^{\min(n,k)} \frac{(-1)^s}{4^s s!(k-s)!(n-s)!} \int_0^{+\infty} e^{-2\varepsilon} \varepsilon^{\frac{n+k+l-2s}{2}} d\varepsilon.$$

(3.13)

Let us denote $\dfrac{\varpi_{n,k}+l}{2} = \lambda \in \mathbb{Z}$, then $l = 2\lambda - \varpi_{n,k}$ and

$$\frac{n+k+l-2s}{2} = \frac{n+k+2\lambda-\varpi_{n,k}-2s}{2} = k+\lambda-s.$$

Consequently,

$$\int_0^{+\infty} e^{-2\varepsilon} \varepsilon^{\frac{n+k+l-2s}{2}} d\varepsilon = \int_0^{+\infty} e^{-2\varepsilon} \varepsilon^{k+\lambda-s} d\varepsilon = \frac{1}{2^{k+\lambda-s+1}} \int_0^{+\infty} e^{-\tau} \tau^{k+\lambda-s} d\tau = \frac{1}{2^{k+\lambda-s+1}} \Gamma(k+\lambda-s+1),$$ (3.14)

where $\tau = 2\varepsilon$ and considered that $\int_0^{+\infty} e^{-\tau} \tau^{z-1} d\tau = \Gamma(z)$, $z = k+\lambda-s+1$. In the view of (3.14), the integral (3.13) is of the form

$$\int_0^{+\infty} e^{-2\varepsilon} \varepsilon^{\frac{l}{2}} \Upsilon_{n,k}\left(\sqrt{2\varepsilon}\right) d\varepsilon = 2^{\frac{n+k-l}{2}-1} \sqrt{n!k!} \sum_{s=0}^{\min(n,k)} \frac{(-1)^s}{2^s s!(k-s)!(n-s)!} \Gamma\left(\frac{k+n+l}{2}-s+1\right).$$ (3.15)

We substitute the expressions (3.15) and (3.11) into the initial integral (3.9) and obtain

$$\int_0^{2\pi} d\varphi \int_0^{+\infty} W \delta U d\varepsilon =$$

$$= \frac{1}{\pi\hbar} \sum_{n,k=0}^{+\infty} \left|\rho_{k,n}\right| \vec{n}_k^T \vec{n}_n \sum_{l=0}^{+\infty} a_l \left(\frac{2\hbar}{m\omega}\right)^{1/2} 2^{\frac{n+k-l}{2}-1} \sqrt{n!k!} \times$$ (3.16)

$$\times \sum_{s=0}^{\min(n,k)} \frac{(-1)^s}{2^s s!(k-s)!(n-s)!} \Gamma\left(\frac{k+n+l}{2}-s+1\right) \begin{cases} \dfrac{2\pi}{2^l} C_l^{\frac{n-k+l}{2}}, & \text{if } \dfrac{n-k+l}{2} \in \mathbb{Z}, \ |n-k| \le l \\ 0, & \text{otherwise.} \end{cases}$$

or

$$I_2 = \hbar \int_0^{2\pi} d\varphi \int_0^{+\infty} W \delta U d\varepsilon = \sum_{n,k=0}^{+\infty} \left|\rho_{k,n}\right| \cos(\alpha_k - \alpha_n) \sqrt{2^{n+k} n!k!} \times$$

$$\times \sum_{\frac{n-k+l}{2} \in \mathbb{Z},\ l \ge |n-k|}^{+\infty} a_l \left(\frac{\hbar}{4m\omega}\right)^{1/2} C_l^{\frac{n-k+l}{2}} \sum_{s=0}^{\min(n,k)} \frac{(-1)^s \left(\dfrac{k+n+l}{2}-s\right)!}{2^s s!(k-s)!(n-s)!}.$$ (3.17)

Substitution of the integrals (3.7) and (3.17) into the expression (3.5) gives the expression (3.3). Theorem 8 is proved.

**Remark**

The expression (3.3) for the energy $\langle\langle\mathcal{E}\rangle\rangle$ of the quantum system consists of two main summands. The first summand corresponds to the energy associated with the diagonal elements of the universal density matrix $\mathcal{W}$. Therefore, the «projections» of the states of an arbitrary quantum system on the oscillator states give a superposition of the energy levels $\langle\langle\varepsilon_n\rangle\rangle$ of the harmonic oscillator (3.3). The second summand in the expression (3.3) contains the energy associated with the off-diagonal elements of the universal density matrix $\mathcal{W}$. This energy is a superposition of the energies of the «mixed» states associated with the difference between an arbitrary quantum system and the harmonic oscillator.

Let us consider an example of a quantum system with a potential

$$U(x) = \frac{m\omega^2 x^2}{2} + \mu x^4. \tag{3.18}$$

The first summand in the potential energy corresponds to the potential of the harmonic oscillator. The second summand $\mu x^4$ introduces anharmonicity. It follows from Theorem 8 that all the coefficients $a_l$ will equal zero, except for the coefficient $a_4 = \mu$. Consequently, when calculating the sum over $l$ (3.3), a condition is imposed on $n$ and $k$

$$\frac{\varpi_{n,k}}{2} + 2 \in \mathbb{Z} \Rightarrow \varpi_{n,k} = 2j, \ j \in \mathbb{Z}, \tag{3.19}$$
$$|\varpi_{n,k}| \le l = 4 \Rightarrow j = 0, \pm 1, \pm 2, \ \varpi_{n,k} = -4, -2, 0, 2, 4.$$

From (3.19) it follows that only five diagonals will be taken from the universal density matrix $\mathcal{W}$: the main diagonal $\varpi_{n,n} = 0$; the second $\varpi_{n,n+2} = -2$ and fourth $\varpi_{n,n+4} = -4$ above; the second $\varpi_{n+2,n} = 2$ and fourth $\varpi_{n+4,n} = 4$ below. The remaining elements of the matrix $\mathcal{W}$ will not be used when calculating the energy $\langle\langle\mathcal{E}\rangle\rangle$ for a quantum system with the potential (3.18).

**Conclusion**

In this paper, a new representation of the Wigner function is proposed through a universal density matrix $\mathcal{W}$ in the phase space. In §1, the explicit form of the matrix $\mathcal{W}$ has been obtained and its properties have been investigated. In parallel with this result, new polynomials $\mathcal{P}_{n,k}(z_1, z_2)$, $\Upsilon_{n,k}(x)$ have been introduced in §1, and their properties, such as orthogonality and association with the Laguerre polynomials, have been investigated. With the help of the new polynomials, the universal density matrix $\mathcal{W}$ is constructed.

Obtaining an explicit form (1.43) for the Wigner function has made it possible to analyze the general character of the behavior of the function $W$. Many papers discuss the dissipative nature of the solution in the phase space [40–42]. Of course, the causes of dissipation underpin the phenomenological construction of the Wigner function (i.1). Analyzing the structure of the universal density matrix $\mathcal{W}$, one can consider the presence of dissipations as the presence of off-diagonal elements in the matrix. Thus, the oscillations of off-diagonal elements can be considered as dissipations.

At the macro level, when averaging over the space of momentum (velocities), the effect of dissipations goes away. So in §2 (Property 5), the mean value of the acceleration flux $\left\langle\left\langle \dot{\vec{v}} \right\rangle\right\rangle$, in contrast to $\left\langle \dot{\vec{v}} \right\rangle$, no longer contains direct information about the microworld. The energy $\left\langle\left\langle \mathcal{E} \right\rangle\right\rangle$ of such a system can be represented as a superposition of the energies (3.3). Depending on the type of the potential $U$ (3.18), not all elements of the matrix $\mathcal{W}$ (3.19) will participate in the superposition (3.3).

**Acknowledgements**

This work was supported by the RFBR No. 18-29-10014.

**References**

1. P. A. M. Dirac, The Principles of Quantum Mechanics (Oxford University Press, New York, 1930).
2. A. Bohrn, Principles of Quantum Mechanics: Fundamentals and Applications (Springer, New York, 1979).
3. L.D. Landau, E.M. Lifshitz, Quantum Mechanics, vol.3, Pergamon Press, 1977, 677 p.
4. E.P. Wigner, On the quantum correction for thermodynamic equilibrium, Phys. Rev. 40 (June 1932) 749—759
5. H. Weyl, The Theory of Groups and Quantum Mechanics (Dover, New York, 1931).
6. M. Hillery, R. F. O'Connell, M. O. Scully, and E. P. Wigner, Phys. Reps. 106, 121 (1984)
7. Agarwal G.S., Wolf E. Calculus for Functions of Noncommuting Operators and General Phase-Space Methods in Quantum Mechanics. II. Quantum Mechanics in Phase Space // Phys. Rev. D. 1970. V. 2. P. 2187-2205.
8. Agarwal G.S., Wolf E. Quantum Dynamics in Phase Space // Phys. Rev. Lett. 1968. V. 21. P. 180-183.
9. Husimi K. Some Formal Properties of the Density Matrix // Proc. Phys. Math. Soc. Jpn. 1940. V. 22. P. 264-314.
10. Kano Y. A new phase-space distribution function in the statistical theory of the electromagnetic field // J. Math. Phys. 1965. V. 6. P. 1913-1915.
11. Glauber R.J. Photon correlations // Phys. Rev. Lett. 1963. V. 10. P. 84-86.
12. Sudarshan E.C.G. Equivalence of semiclassical and quantum mechanical descriptions of statistical light beams // Phys. Rev. Lett. 1963. V. 10. P. 277-279.
13. Cahill K.E., Glauber R.J. Density Operators and Quasiprobability Distributions // Phys. Rev. A. 1969. V. 177. P. 1882-1902.
14. G. B. Folland. Harmonic Analysis in Phase space. Annals of Mathematics studies, Princeton University Press, Princeton, N.J., 1989.
15. R. G. Littlejohn. The semiclassical evolution of wave packets. Physics Reports 138( 4-5):193- 291, 1986
16. M. W. Wong. Weyl Transforms. Springer, 1998.
17. E.E. Perepelkin, B.I Sadovnikov, N.G. Inozemtseva, Riemann surface and quantization, Annals of Physics (2016) DOI: 10.1016/j.aop.2016.11.012
18. N. Wiener, Hermitian Polynomials and Fourier Analysis, J. Mathematics and Physics 8 (1929) 70-73.
19. V. Namias, The fractional order Fourier transform and its application to quantum mechanics, J. Inst. Appl. Math. 25, 241–265 (1980).
20. Vlasov A.A., Many-Particle Theory and Its Application to Plasma, New York, Gordon and Breach, 1961, ISBN 0-677-20330-6; ISBN 978-0-677-20330-0
21. Vlasov A.A., Statisticheskie funkcii raspredelenija, Moscow, Nauka, 1966, 356 p.

22. Vlasov A.A., Inozemtseva N.G., The existence of four types of acoustic waves in the static model of a crystal, MSU bulletin, Series 3, 1976, v. 17, №2, p 151-159
23. Vlasov A.A., Inozemtseva N.G., The basic types of acoustic waves transporting acoustic spin in crystals, DAN SSSR, 1975, v. 225, №2, p 276-279
24. Perepelkin E.E., Sadovnikov B.I., Inozemtseva N.G., The properties of the first equation of the Vlasov chain of equations, J. Stat. Mech. (2015) P05019
25. E. E. Perepelkin, B. I. Sadovnikov, N. G. Inozemtseva, The new modified Vlasov equation for the systems with dissipative processes, Journal of Statistical Mechanics: Theory and Experiment, (2017) № 053207
26. J.E. Moyal, Quantum mechanics as a statistical theory, Proceedings of the Cambridge Philosophical Society, 45, 99–124 (1949). doi:10.1017/S0305004100000487
27. Groenewold H.J. On the principles of elementary quantum mechanics // Physica. 1946. V. 12. P. 405-460.
28. W. Koepf, Identities for families of orthogonal polynomials and special functions, Integral Transforms and Special Functions 5, (1997)
29. W. A. Al-Salam, Operational representations for Laguerre and other polynomials, Duke Math (1964) J. 31 (1): 127-142
30. Simpao, Valentino A. (2014). «Real wave function from Generalised Hamiltonian Schrodinger Equation in quantum phase space via HOA (Heaviside Operational Ansatz): exact analytical results». Journal of Mathematical Chemistry. 52 (4): 1137–1155. doi:10.1007/s10910-014-0332-2. ISSN 0259-9791.
31. D. B. Fairliet, C. A. Manoguei, The formulation of quantum mechanics in terms of phase space functions-the third equation, J. Phys. A: Math. Gen. 24 (1991) 3807-3815. Printed in the UK
32. J. J. Wlodarz, On quantuin mechanical phase-space wave functions, J. Chem. Phys. 100 (lo), 15 May 1994
33. Møller, K. B., Jørgensen, T. G., & Torres-Vega, G. (1997). On coherent-state representations of quantum mechanics: Wave mechanics in phase space. Journal of Chemical Physics, 106(17), 7228-7240. DOI: 10.1063/1.473684
34. T. L. Curtright, D. B. Fairlie, C. K. Zachos, A concise treatise on quantum mechanics in phase space, World Scientific Publishing 2014, ISBN 978-981-4520-43-0
35. E.E. Perepelkin, B.I Sadovnikov, N.G. Inozemtseva, The quantum mechanics of high-order kinematic values, Annals of Physics (2019) vol. 401 pp. 59–90
36. D. Bohm, B.J. Hiley, and P.N. Kaloyerou, «An ontological basis for the quantum theory», Phys. Rep. 144, 321-375 (1987)
37. D. Bohm and B.J. Hiley, The Undivided Universe: An Ontological Interpretation of Quantum Theory (Routledge, London, 1993)
38. D. Bohm, Phys. Rev. 85, 166-193 (1952).
39. L. de Broglie, Une interpretation causale et non lineaire de la mecanique ondulatoire: la theorie de la double solution, Gauthiers-Villiars, Paris (1956). English translation: Elsevier, Amsterdam (1960).
40. D. Kakofengitis, O. Steuernagel, Wigner's quantum phase-space current in weakly-anharmonic weakly-excited two-state systems, Eur. Phys. J. Plus (2017) 132: 381
41. T. L. Curtright, D. B. Fairlie, C. K. Zachos, A concise treatise on quantum mechanics in phase space, World Scientific Publishing 2014, ISBN 978-981-4520-43-0
42. Møller, K. B., Jørgensen, T. G., & Torres-Vega, G. (1997). On coherent-state representations of quantum mechanics: Wave mechanics in phase space. Journal of Chemical Physics, 106(17), 7228-7240.